\documentclass[manuscript,nonacm]{acmart}
\usepackage[T2A]{fontenc}
\AtBeginDocument{%
  }

\usepackage{subcaption} 

\setcopyright{acmlicensed}
\copyrightyear{2018}
\acmYear{2018}
\acmDOI{XXXXXXX.XXXXXXX}
\acmConference[Conference acronym 'XX]{Make sure to enter the correct
  conference title from your rights confirmation email}{June 03--05,
  2018}{Woodstock, NY}
\acmISBN{978-1-4503-XXXX-X/2018/06}

\begin{document}

\title{From Forgeries to Foundation Models: A Systematic Survey of Identity Document Attack and Detection}




 \author{Gourab Das}
 \orcid{0009-0009-6804-7255}
\email{gourab.das@iiitdwd.ac.in}
 \affiliation{%
   \institution{Indian Institute of Information Technology Dharwad (IIIT Dharwad)}
   \country{India}
}

 \author{Pavan Kumar C}
 \orcid{0000-0002-4907-5412}
 \email{pavan@iiitdwd.ac.in}
 \affiliation{%
   \institution{Indian Institute of Information Technology Dharwad (IIIT Dharwad)}
   \country{India}
 }

 \author{Raghavendra Ramachandra}
 \orcid{0000-0003-0484-3956}
 \email{raghavendra.ramachandra@ntnu.no}
 \affiliation{%
  \institution{SAFE Center, Norwegian University of Science and Technology (NTNU)}
  \city{Gjøvik}
   \country{ Norway}
 }

\renewcommand{\shortauthors}{Das et al.}

\begin{abstract}

Identity document forgery has undergone a fundamental capability shift:
generative AI tools now enable high-fidelity document synthesis and
field-level manipulation with minimal technical expertise, while detection
methods remain constrained by benchmarks that do not reflect this threat.
The resulting attack surface spans physical presentation, digital
injection, and fully generative synthesis, introducing distinct forensic
failure modes that require a unified threat model and evaluation
framework. This survey provides, to our knowledge, the first unified
treatment of Presentation Attacks, Digital Injection Attacks, and
GenAI-driven synthesis within a single identity verification threat model.
We trace detection methodologies from rule-based heuristics through
forensic localisation, injection-aware pipelines, foundation models, and
few-shot frameworks. A systematic audit of public datasets from
2019--2025 exposes a persistent Reality Gap between benchmark conditions
and operational deployment. We further analyse large multimodal models for
identity document manipulation, identifying Script-Dependent Generative
Instability (SDGI) as a recurring typographic failure mode in non-Latin
script inpainting. Finally, zero-shot benchmarking on unseen synthesised
ID cards shows that even the strongest publicly available models achieve
APCER values above 25\% under security-oriented operating conditions,
highlighting substantial limits in cross-domain generalisation. We
conclude by outlining future directions toward forensically grounded,
privacy-preserving, and legally accountable identity verification systems.

\end{abstract}



\received{XXXXXXXXXXXX}
\received[revised]{XXXXXXXXXXXX}
\received[accepted]{XXXXXXXXXXXX}

\maketitle

\section{Introduction}

Identity verification has emerged as a fundamental pillar of contemporary digital society, underpinning critical functions in financial services, border control, public welfare systems, and online service provisioning. The global transition from physical to camera-based identity card verification has enabled remote identity verification at unprecedented scale---but has simultaneously exposed serious structural vulnerabilities. Readily available image editing tools and, more recently, generative AI have fundamentally altered the threat landscape, enabling attackers with minimal technical expertise to create visually convincing fake documents. These developments have elevated identity document forgery from an isolated criminal activity into a systemic, scalable threat.

\textbf{Real-world motivation and threat landscape:} Recent incidents illustrate the operational relevance of identity document forgery and the breadth of attack surfaces. Documented cases include (i) AI-generated documents bypassing automated KYC verification, (ii) forged sovereign identity documents produced without technical expertise, (iii) large-scale synthetic identity fraud causing systemic financial losses, and (iv) formal law enforcement and regulatory alerts confirming active criminal exploitation. Representative public reports include:

\begin{itemize}

\item \textbf{2024--2026 (Global/USA, \textit{OnlyFake}):} An underground platform sold photorealistic fake passports and driver's licences for roughly \$15 each, capable of bypassing KYC checks on major platforms; its operator was later arrested and pleaded guilty after selling more than 10,000 AI-generated documents worldwide~\cite{cox2024onlyfake,frankonfraud2026onlyfake,doj2026onlyfake,biometricupdate2026onlyfake}.

\item \textbf{2024 (USA, FinCEN/FBI):} FinCEN and the FBI issued alerts confirming criminals were using generative AI to manipulate or fabricate identity documents (e.g., driver's licences) for bank, check, and loan fraud, and for impersonation~\cite{fincen2024, fbi2024}.

\item \textbf{2025 (India, Bengaluru):} A public demonstration showed that a generative AI model could produce highly realistic Aadhaar and PAN card replicas, exposing the limits of visual inspection-based verification~\cite{ndtv2025nanobana, hindustantimes2025nanobana}.

\item \textbf{2025 (South Korea):} The North Korean group \textit{Kimsuky} reportedly used generative AI to forge South Korean military ID cards in spear-phishing campaigns against defense-sector personnel~\cite{genians2025kimsuky, bloomberg2025kimsuky}.

\item \textbf{2025 (India):} A Forrester--Experian survey found that 69\% of organizations considered their KYC infrastructure inadequate against AI-generated identity documents~\cite{forrester2025india}.

\item \textbf{2023--2025 (USA/Global, financial impact):} The aggregate scale is substantial: US identity fraud cost approximately \$27 billion across 18 million victims in 2024~\cite{javelin2025}; AI-related cybercrime complaints exceeded 22,000 with losses over \$800 million in 2025~\cite{fbi2025aiscams}; and global synthetic identity fraud surpassed \$35 billion annually~\cite{fedbos2025}.

\end{itemize}

While such reports vary in evidential detail and technical disclosure, collectively they motivate robust identity document forgery detection methods and, more broadly, secure verification system design that does not rely on visual inspection alone as a high-assurance signal~\cite{fincen2024, iproov2024}.

\begin{figure}[t]
\centering
\includegraphics[width=\columnwidth]{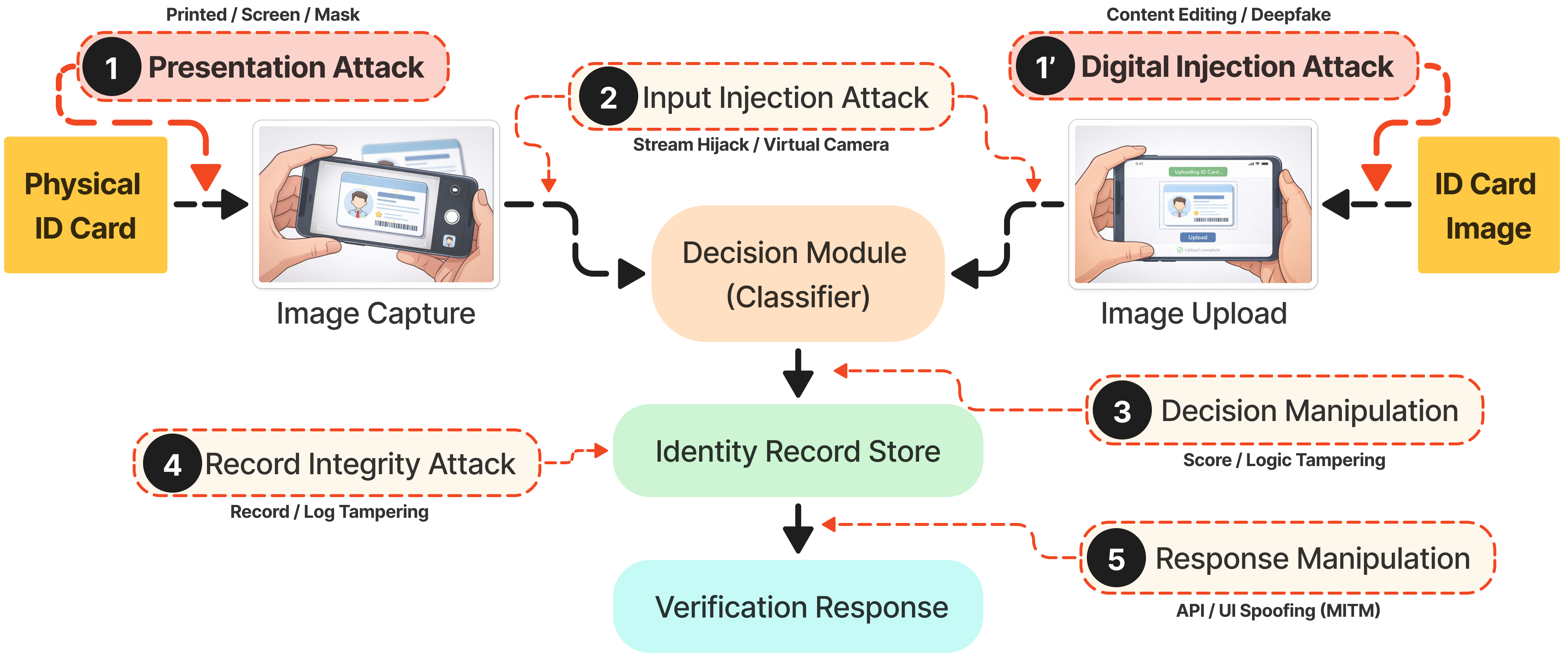}
\caption{Attack surface in identity document verification pipeline. Attacks occur at five stages: (1) Presentation Attack using printed/screen/manipulated documents at capture, (1') Digital Injection Attack through content editing post-capture, (2) Input Injection Attack via stream hijacking, (3) Decision Manipulation through score tampering, (4) Record Integrity Attack via log tampering, and (5) Response Manipulation through API spoofing.}
\label{fig:attack_surface}
\end{figure}

Identity card verification operates through multiple stages, as
illustrated in Figure~\ref{fig:attack_surface}. This survey addresses
two attack vectors that target identity documents at different points
in this pipeline: \textbf{Presentation Attacks (PA)}, where a
physically falsified document is presented to a capture device, and
\textbf{Digital Injection Attacks (IA)}, where manipulated imagery is
submitted into the pipeline without physical capture. Their precise
definitions and attack-surface topology are established in
Section~\ref{sec:attack_surface}.

\begin{figure}[tb]
\centering
\includegraphics[width=\columnwidth]{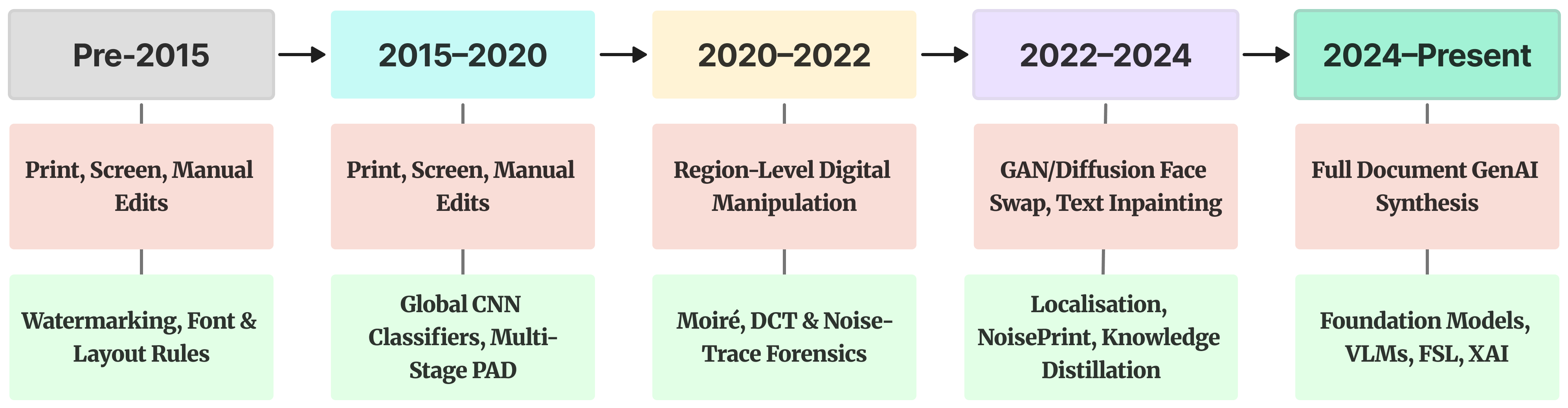}
\caption{Co-evolution of attack capability and detection methodology in
identity document forgery, from physical presentation attacks to
AI-assisted manipulation and generative document synthesis.}
\label{fig:research_evolution}
\end{figure}

\textbf{Research evolution.}
Identity document forgery and its detection have followed a
co-evolutionary trajectory in which advances in attack capability
consistently triggered methodological responses on the detection side.
Figure~\ref{fig:research_evolution} summarises this progression.

\begin{itemize}
    \item \textbf{Pre-2015 (Physical forgeries --- rule-based detection):}
    Attacks were largely limited to print reproductions, screen
    recaptures, and manual text alteration, while detection relied on
    watermark inspection, font analysis, and layout consistency checks.

    \item \textbf{2015--2020 (Composite forgeries --- global classification):}
    Consumer editing tools enabled copy-move and composite document
    manipulations, motivating CNN-based classifiers that performed well
    in controlled settings but generalised poorly across datasets and
    attack types.

    \item \textbf{2020--2022 (Recapture and region edits --- forensic
    micro-artefact analysis):} As region-level edits became more common,
    detection shifted toward frequency-domain and noise-trace cues such
    as moir\'{e} patterns, DCT artefacts, and recapture signatures.

    \item \textbf{2022--2024 (AI-assisted manipulation --- localisation and
    semantic reasoning):} GAN- and diffusion-based face and text editing
    weakened traditional forensic cues, leading to increased emphasis on
    localisation, device fingerprints, and semantic inconsistency
    analysis.

    \item \textbf{2024--Present (GenAI synthesis --- foundation models and
    few-shot learning):} Large multimodal models enabled increasingly
    realistic field-level and full-document synthesis, prompting a shift
    toward foundation-model-based detection, few-shot learning, and
    explainable analysis, while robust generalisation remains unresolved.
\end{itemize}

\subsection{Literature Collection and Screening Protocol}
\label{sec:protocol}

We assembled a corpus of publications on identity document forgery and
detection across three attack classes: physical presentation attacks,
digital injection attacks, and generative-AI-driven synthesis. The primary
corpus is restricted to works in which identity documents are the primary
or co-primary experimental object. General image forensics, biometric
anti-spoofing, OCR, and document-understanding papers are therefore
excluded unless they provide the direct technical basis for an
ID-document-specific method; such works are treated as methodological
references rather than members of the primary corpus. Coverage spans
2015--2026, with selected pre-2015 works retained only where they provide
foundational context for identity document security. The final search pass
was completed on 10 June 2026.

\textbf{Search strategy.}
Records were retrieved from Scopus, IEEE Xplore, ACM Digital Library, and
SpringerLink, and supplemented by a depth-limited Google Scholar search.
We also manually examined proceedings of IJCB, IEEE FG, and BTAS, where
identity document PAD results are often reported before journal archiving.
The Boolean query was:

\smallskip
\noindent\texttt{\small
("presentation attack detection" OR "anti-spoofing" \\
 OR "digital injection attack" OR "document forgery" \\
 OR "document fraud" OR "identity fraud" \\
 OR "morphing attack" OR "face photo substitution" \\
 OR "document authentication" OR "print-scan" \\
 OR "e-passport" OR ePassport OR eID \\
 OR "generative AI" OR GenAI OR deepfake* \\
 OR "synthetic image*" OR "AI-generated") \\
AND ("identity document*" OR "identity card*" \\
 OR "ID card*" OR passport* OR "national identity")}
\smallskip

\noindent
The query was intentionally broad to maximise recall. Consequently, many
records retrieved from adjacent areas such as general deepfake detection,
face anti-spoofing, and image manipulation forensics were removed during
screening because they did not evaluate identity documents as a dedicated
security object. The Google Scholar query returned 1668 records; we
screened the first 1{,}000 results, following the reliability boundary
discussed by Gusenbauer and Haddaway~\cite{gusenbauer2020academic}.

\begin{figure*}[t]
\centering
\includegraphics[width=\textwidth]{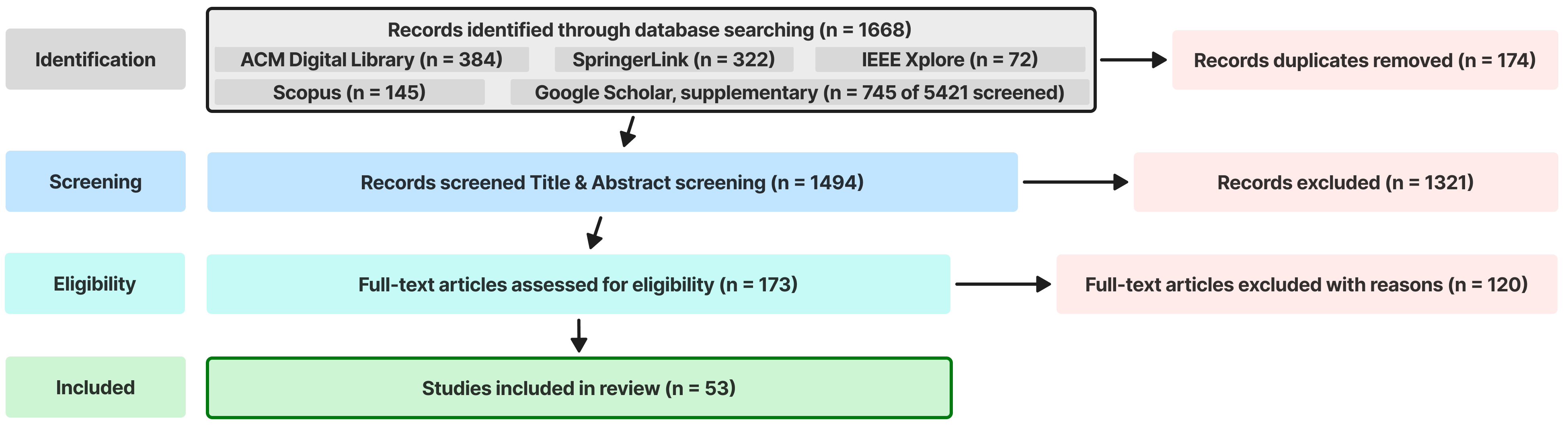}
\caption{PRISMA 2020 flow diagram for literature identification and
screening. The primary corpus contains publications in which identity
document security is the primary or co-primary experimental focus;
methodological references from adjacent fields are tracked separately and
identified at first citation.}
\label{fig:prisma_flow}
\end{figure*}

\textbf{Screening and inclusion criteria.}
Screening was conducted in two stages: title-and-abstract review followed
by full-text assessment. A publication was included in the primary corpus
if it made a substantive contribution to identity document forgery or
detection through at least one of the following: (a) a detection
architecture evaluated on identity documents; (b) a forensic method for
characterising or localising forgery artefacts in document images; (c) an
ID-document-specific dataset, benchmark, or evaluation protocol; (d) a
generative or synthetic attack pipeline targeting identity documents; or
(e) a security analysis of identity verification pipelines in which the
document image is a central attack surface.

We excluded studies that met any of the following criteria:
\begin{enumerate}
  \renewcommand{\labelenumi}{(\roman{enumi})}
  \item identity documents were not the primary contribution or a directly
  evaluated experimental component, including works on general image
  forgery, face liveness, fingerprint spoofing, or deepfake detection
  without identity document experiments;
  \item the work addressed OCR, layout analysis, information extraction,
  or document recognition without a forgery, authentication, or security
  component;
  \item the study lacked sufficient methodological or experimental detail
  to assess its evaluation protocol or performance claims.
\end{enumerate}
When both a preprint and a peer-reviewed version of the same work were
available, the more complete version was retained. The selection process
followed PRISMA 2020 guidelines~\cite{page2021prisma} and is summarised
in Figure~\ref{fig:prisma_flow}.

\begin{figure}[tb]
\centering
\includegraphics[width=\columnwidth]{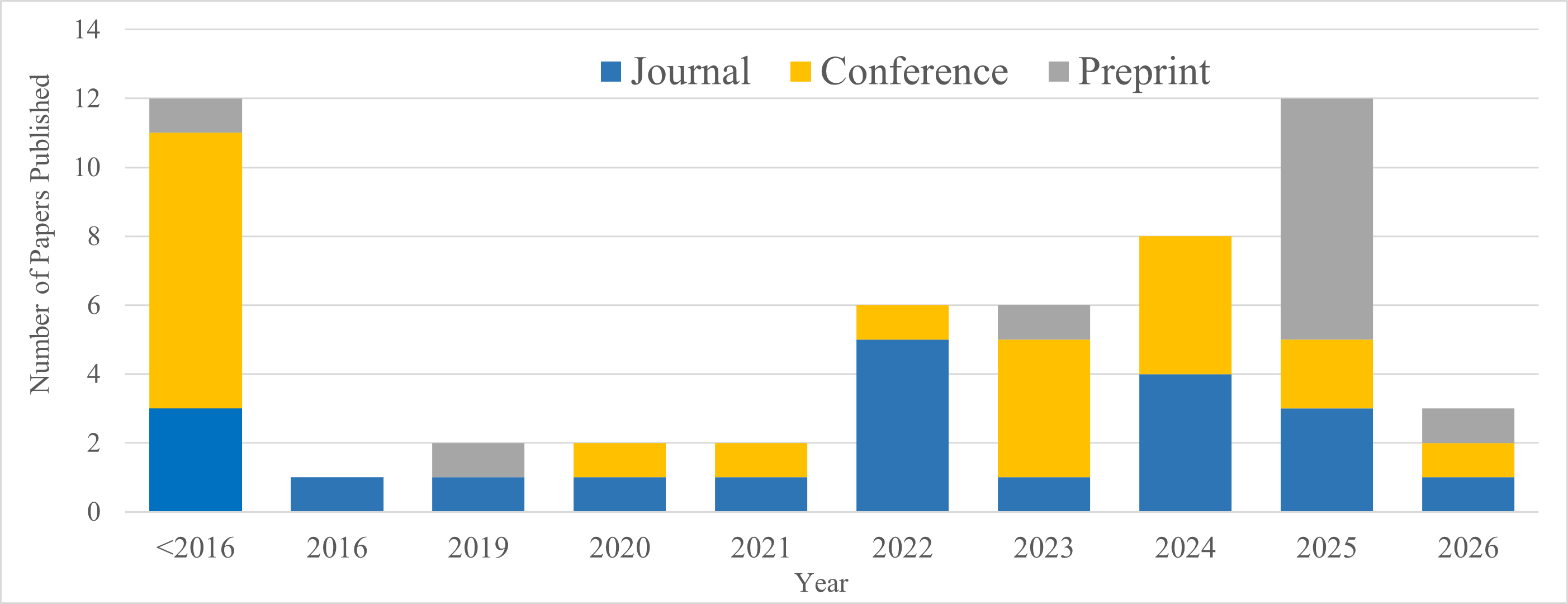}
\caption{Year-wise distribution of the 53 primary-corpus publications
retained after PRISMA 2020 screening. The increase after 2022 coincides
with the emergence of ID-card-specific PAD benchmarks and the rapid
diffusion of generative AI tools.}
\label{fig:year_wise}
\end{figure}

\textbf{Corpus statistics and coverage.}
Of the 1{,}668 records identified, 174 duplicates were removed and 1{,}321 records were excluded at title-and-abstract screening. Full-text assessment of the remaining 173 articles excluded a further 120 on grounds~(i)--(iii), yielding \textbf{53 publications} in the primary corpus. To our knowledge, this corpus accounts for the substantial majority
of English-language publications in which identity document security constitutes the primary or co-primary experimental focus. This corpus size reflects both the focused scope and recency of the field: most public ID-card PAD benchmarks have appeared since 2019, while dedicated work on generative identity document attacks
has grown mainly after 2022. The year-wise distribution of the retained publications is shown
in Figure~\ref{fig:year_wise}.

\subsection{Research Questions}

This survey is guided by four research questions defined prior to the
literature review:

\paragraph{RQ1 (Threat Landscape and Taxonomy)}
How has identity document forgery evolved from opportunistic physical presentation attacks to AI-assisted manipulation and generative synthesis, and how do these attack classes differ in attacker capability, forensic traceability, and detection difficulty?

\paragraph{RQ2 (Detection Methods)}
How have detection methods progressed from rule-based heuristics to deep learning, forensic localisation, injection-aware architectures, foundation models, and few-shot frameworks; and what limitations persist at each stage?

\paragraph{RQ3 (Datasets, Benchmarks, and the Reality Gap)}
To what extent do existing datasets and evaluation protocols capture real-world forgery complexity, and which gaps in attack coverage, localisation ground truth, and GenAI representation contribute to the Reality Gap between benchmark performance and operational deployment?

\paragraph{RQ4 (Generative AI Capabilities and Detection Opportunities)}
What capabilities and structural limitations do large multimodal generative models exhibit in identity document manipulation, and which typographic, forensic, or layout-level artefacts remain exploitable for detection?


\subsection{How This Study Differs from Existing Work}

Prior surveys have examined presentation attack detection for individual biometric modalities and, more recently, image manipulation detection in modality-agnostic settings. Sousedik and Busch~\cite{sousedik2014presentation}
reviewed hardware- and software-based liveness detection for fingerprint sensors; Ramachandra et al.\cite{ramachandra2017presentation} surveyed 2D and 3D face presentation attacks and countermeasures; and Pooshideh
et al.~\cite{pooshideh2024presentation} updated this line of work with handcrafted and deep-learning face anti-spoofing methods from 2004--2024. Zanardelli et al.~\cite{zanardelli2023image} surveyed deep-learning-based image manipulation and forgery detection more broadly. These surveys are important methodological foundations, but they do not treat identity documents as a distinct security object combining layout-constrained text, face-photo regions, substrate cues, printing artefacts, and document-level security features.

The closest prior work is the ID-card PAD survey of Ruiz et al.~\cite{ruiz2025identity}. However, its scope is restricted to
physical presentation attacks, excluding digital injection attacks and GenAI-driven full-document synthesis. It also reports results from the surveyed literature rather than evaluating detectors and foundation models under a common protocol. In contrast, this survey unifies physical, digital, and generative identity document attacks; analyses datasets and benchmarks through the lens of the Reality Gap; and includes an independent zero-shot evaluation of contemporary multimodal foundation models on a shared evaluation set.

\subsection{Scope and Contributions of This Survey}
\label{sec:scope}

This survey covers identity document forgery and detection across three
attack classes: physical presentation attacks, digital injection attacks,
and generative-AI-driven document synthesis. It reviews methods ranging
from rule-based heuristics and forensic analysis to deep learning,
foundation models, and few-shot detection. Related biometric PAD
literature is discussed only where it directly informs document-level
detection.

The primary contributions are:
\begin{itemize}
    \item A unified threat framework that places presentation attacks,
    digital injection attacks, and GenAI-driven synthesis within a single
    identity verification lifecycle.
    \item A capability-driven forgery taxonomy covering four classes of
    attacks, from opportunistic physical manipulation to fully generative
    document synthesis.
    \item A systematic audit of public ID-card forgery datasets and
    evaluation protocols from 2019--2025, exposing the Reality Gap between
    benchmark conditions and operational deployment.
    \item An empirical analysis of large multimodal generative models for
    identity document manipulation, including Script-Dependent Generative
    Instability (SDGI) in non-Latin script inpainting.
    \item A zero-shot benchmarking study on unseen synthetic ID cards,
    providing a common reference point for cross-domain generalisation
    under operationally realistic constraints.
    \item A consolidation of open challenges and future directions spanning
    benchmark design, cryptographic trust signals, explainability, and
    privacy-preserving continual learning.
\end{itemize}

\textbf{Organisation of the article:} Section~\ref{sec:lifecycle} establishes the verification lifecycle and attack surface topology. Section~\ref{sec:taxonomy} presents the capability-driven forgery taxonomy. Section~\ref{sec:detection} surveys detection methods from rule-based heuristics to foundation models. Section~\ref{sec:benchmarks} audits public datasets and evaluation protocols. Section~\ref{sec:genai} examines generative AI capabilities and forensic implications. Section~\ref{sec:ethics} addresses deployment constraints and ethical considerations. Section~\ref{sec:future} outlines future directions, and Section~\ref{sec:conclusion} concludes.

\section{Verification Lifecycle, Attack Surfaces, and Forensic Properties}
\label{sec:lifecycle}

Identity document verification is a multi-stage pipeline in which trust established at issuance is propagated and verified through successive stages up to automated verification and human review.
Unlike natural images or generic document scans, identity documents are
purpose-designed security artefacts whose layout, materials, typography,
and issuance procedures constrain both adversarial manipulation and
forensic detectability. This section defines the verification lifecycle,
acquisition modalities, attack surfaces, and forensic properties that
motivate ID-document-specific detection methods~\cite{centeno2019identity}.

\subsection{End-to-End Identity Verification Lifecycle}
\begin{figure}[tb]
\centering
\includegraphics[width=\columnwidth]{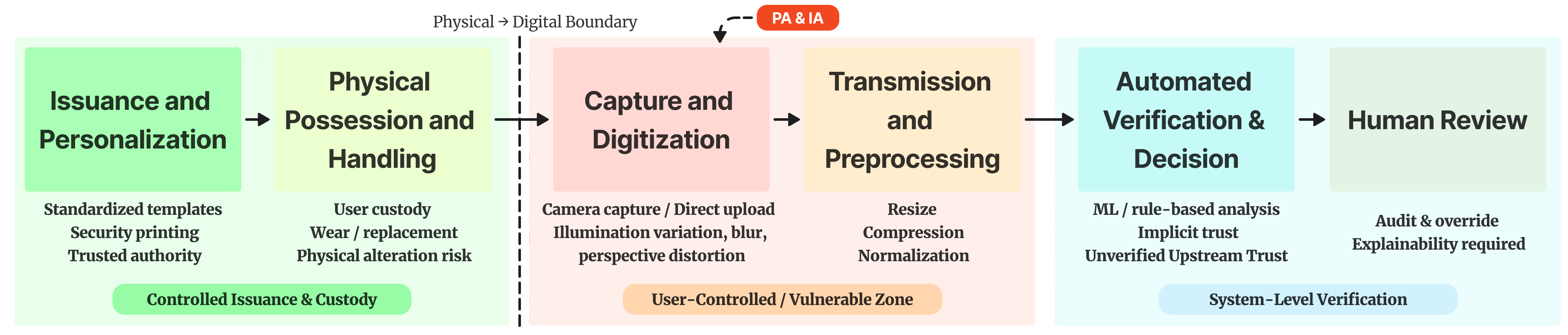}
\caption{Identity document lifecycle showing trust assumptions and
vulnerability profiles from issuance to automated and human verification.}
\label{fig:lifecycle}
\end{figure}

An identity document progresses through a sequence of stages, each with distinct trust assumptions and vulnerability profiles
(Figure~\ref{fig:lifecycle}). \textit{Issuance and personalisation} are performed by trusted authorities using standardised templates, regulated fonts, and controlled printing processes, making compromise at this stage rare but potentially systemic~\cite{ICAO9303P2_2021}. During \textit{physical possession and handling}, documents may be worn, lost,
replaced, or deliberately altered, enabling partial substitutions and overlays~\cite{centeno2019identity}. \textit{Capture and digitisation} introduce illumination variation, blur, perspective distortion, sensor noise, and compression artefacts that may either reveal or obscure manipulation traces~\cite{piva2013overview}. In \textit{transmission and
preprocessing}, resizing, normalisation, and recompression may further alter forensic cues~\cite{stamm2013information}. Finally, \textit{automated verification} and, where required, \textit{human-in-the-loop review} must decide authenticity under constraints of accuracy, interpretability, auditability, and evidentiary clarity~\cite{doshi2017towards}.

\subsection{Acquisition and Representation Modalities}
Forensic evidence depends strongly on acquisition modality. Physical
inspection, scanners, mobile cameras, video streams, and chip-based
authentication preserve different subsets of information and therefore
support different detection strategies. Table~\ref{tab:modalities}
summarises their main forensic advantages and limitations.

\begin{table*}[tb]
\centering
\caption{Comparison of identity document acquisition modalities and their forensic implications.}
\label{tab:modalities}
\small
\begin{tabular}{p{2.0cm}p{3.0cm}p{3.8cm}p{4.5cm}}
\toprule
\textbf{Modality} & \textbf{Key Characteristics} & \textbf{Forensic Advantages} & \textbf{Limitations} \\
\midrule
Physical Inspection~\cite{centeno2019identity,ICAO9303P2_2021}
& Direct examination with specialised lighting
& UV-reactive inks, tactile elements, material composition analysis
& Examiner variability; not scalable for remote verification \\
\midrule
Flatbed Scanner~\cite{clarkson2009fingerprinting,5159455}
& High resolution (300--1200 DPI), controlled illumination
& Preserves microprinting, security patterns, geometric consistency, and scanner-specific noise
& High-quality templates may support sophisticated forgery; requires specialised equipment \\
\midrule
Mobile Camera~\cite{piva2013overview}
& Ubiquitous remote capture with automatic processing pipelines
& Sensor noise analysis, moir\'{e} detection, and video-based liveness cues
& Perspective distortion, variable illumination, motion blur, and compression artefacts \\
\midrule
Video Stream~\cite{rossler2019faceforensics++,bakas2021detection}
& Temporal sequence of frames
& Frame consistency, motion cues, and depth cues for injection detection
& Video compression introduces temporal and spatial artefacts \\
\midrule
RFID/NFC Chip~\cite{ICAO_DTC_2024,madlmayr2008nfc,francis2011practical} & Cryptographic authentication and secure element storage & Resistant to visual forgery and supports biometric binding & Requires specialised readers; vulnerable to relay-based presentation attacks even with genuine chips; unsuitable for many remote workflows \\
\bottomrule
\end{tabular}
\end{table*}

Each modality preserves different forensic cues. High-resolution scans
retain microstructure but lack temporal evidence; mobile captures contain
sensor and recapture traces but suffer from uncontrolled imaging
conditions; chip-based checks provide cryptographic assurance but are not
available in many remote verification settings. Effective systems must
therefore be modality-aware rather than assuming that cues reliable in one
acquisition setting will transfer unchanged to another.

\subsection{Attack Surface Mapping: Presentation vs. Injection Attacks}
\label{sec:attack_surface}

As shown in Figure~\ref{fig:attack_surface}, adversaries may introduce
manipulation at several points in the verification pipeline. This survey
focuses on two document-centred attack vectors: \emph{Presentation
Attacks} (PA) and \emph{Injection Attacks} (IA). They differ primarily in
where the forged content enters the system.

\textbf{Presentation Attacks (PA)} occur at the physical capture stage.
The adversary presents a forged, altered, counterfeit, printed, or
screen-displayed document to a legitimate acquisition device such as a
camera or scanner. The system therefore receives a digital capture of a
physically presented artefact~\cite{10744475}.

\textbf{Injection Attacks (IA)} occur in the digital domain, bypassing or replacing the trusted physical capture process entirely. The adversary supplies manipulated or synthetic document imagery through upload interfaces, compromised capture streams, software hooks, or downstream API pathways. Unlike PA, the attack is introduced in the digital domain and may bypass physical substrate, recapture, and sensor-level assumptions ~\cite{korshunov2025fantasyid}.

This distinction separates attacks that depend on physical-world presentation from those that directly target the digital representation or pipeline. It also explains why detectors based on substrate, print-scan, or recapture traces may fail against fully digital or generative attacks. This distinction forms the basis for the capability-driven taxonomy introduced in Section~\ref{sec:taxonomy} and for the detection analysis developed in later sections.

\subsection{Identity Documents as a Distinct Forensic Problem}

Identity documents differ from natural images and generic document images
in ways that justify treating their forgery detection as a distinct
forensic problem class~\cite{bulatov2022towards}.

\begin{itemize}
    \item They follow rigid, template-constrained layouts with fixed
    spatial relationships among fields, portraits, security features, and
    machine-readable zones.

    \item They combine heterogeneous evidence sources, including face
    photographs, text fields, typography, background patterns, security
    printing, substrate cues, and sometimes chip-based trust signals.

    \item Semantic fields such as name, date of birth, document number,
    expiry date, and nationality exhibit cross-field dependencies;
    multilingual documents add script and transliteration consistency
    constraints.

    \item Legitimate documents from the same type and issuance period
    often exhibit low intra-class variability, enabling anomaly detection
    strategies that are less applicable to unconstrained natural images.

    \item They operate in adversarial and regulated environments where
    attackers adapt to detection methods, while verification errors carry
    legal, financial, and personal consequences.
\end{itemize}

These properties explain why methods developed for natural images, generic document analysis, or isolated biometric PAD often require substantial adaptation before they can provide reliable identity document forgery detection. The following section introduces a capability-driven taxonomy of forgery attacks structured around these forensic properties and the attack surfaces established in Section~\ref{sec:attack_surface}.

\section{Capability-Driven Forgery Taxonomy}
\label{sec:taxonomy}

Identity document forgery has evolved alongside advances in image editing,
automation, and generative modelling. This section presents a taxonomy
organised by attacker capability, operational risk, and verification
impact. Rather than classifying attacks only by manipulation type, we use
a capability-driven framework that reflects how attacks scale with tool
access, document knowledge, and technical expertise. We identify four
classes: opportunistic forgeries, structure-preserving forgeries,
AI-assisted localised forgeries, and GenAI-driven full-document forgeries.
Some attack types, especially text-field manipulation, recur across
multiple classes because their forensic sophistication depends strongly on
the tools used. This taxonomy provides the basis for the detection
challenges analysed in later sections.

\begin{figure}[tb]
\centering
\includegraphics[width=\linewidth]{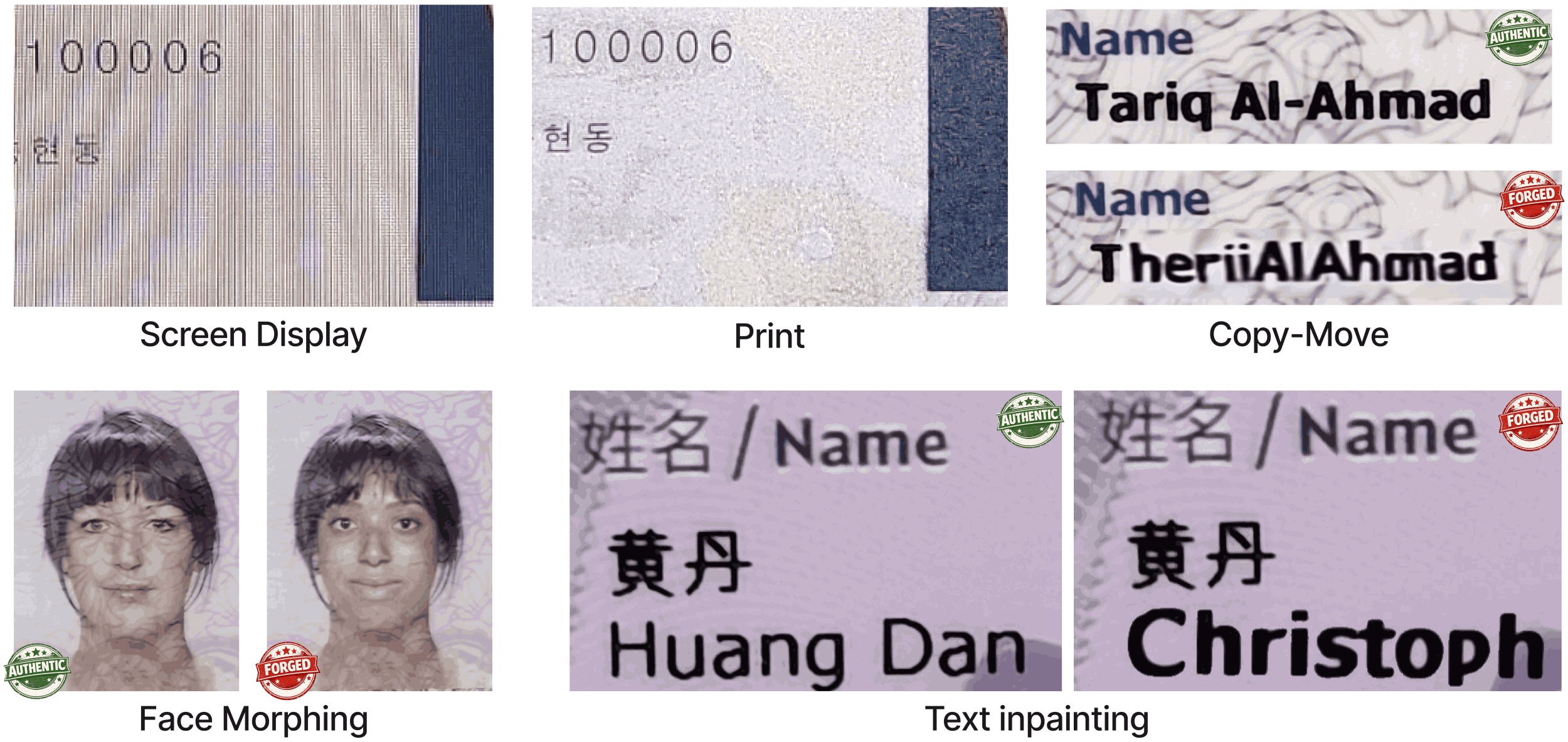}
\caption{Representative examples of presentation and digital injection
attacks on identity documents. Top row: screen-display attack with
moir\'{e} patterns, print attack with halftoning artefacts, and copy-move
forgery with visible font and background inconsistencies. Bottom row:
face morphing and text-field manipulation representative of higher
capability AI-assisted attacks.}
\label{fig:types-of-forgery}
\end{figure}

\subsection{Opportunistic Forgery}

Opportunistic forgeries represent the lowest-capability attack class,
requiring little document knowledge and no specialised generative tools.
This class includes screen attacks, print attacks, and manual text-field
replacement.

Screen attacks occur when a digital image of an identity document is
displayed on an electronic screen and recaptured by the verification
device~\cite{park2023kid34k,polevoy2022document}. More sophisticated
variants use high-pixel-density displays to reduce visible grid patterns
and improve perceptual realism~\cite{munoz2025fakeidet}. Print attacks
involve presenting a printed reproduction of a document, often on glossy
photo paper to approximate the appearance of genuine card substrates
~\cite{tapia2025synid}. Both attack types introduce characteristic
physical artefacts, including moir\'{e} interference from screen recapture
and halftoning structures from printing, which remain detectable by global
classifiers and frequency-domain forensic methods.

Manual text-field replacement at this level involves erasing or covering
a field and inserting replacement content using basic image editing
software~\cite{boned2024synthetic}. Such forgeries often exhibit
incorrect font family or weight, inconsistent spacing and character size,
background discontinuities, and local noise-pattern breaks. These traces
are typically visible to pixel-level or frequency-domain analysis and do
not require generative-model-specific detectors.

\subsection{Structure-Preserving Forgery}

Structure-preserving forgeries assume greater knowledge of document
templates, including font characteristics, field spacing, alignment rules,
and typographic conventions. Rather than synthesising content from
scratch, attackers copy regions such as portraits, signatures, or text
fields from documents of the same template class and transplant them into
a target document while preserving the surrounding layout
~\cite{boned2024synthetic}.

The canonical attack in this class is copy-move forgery. Because copied
regions may originate from genuine documents, typography and local layout
can appear plausible. However, the compositing process may leave edge
discontinuities, blending artefacts, noise-pattern mismatches, and subtle
illumination inconsistencies at region boundaries~\cite{gonzalez2020hybrid}.
If the forged document is reprinted or recaptured, additional print-scan
artefacts may either reveal manipulation or mask weaker splicing cues.

Text manipulation at this level may use semi-automated region filling to
reconstruct local background texture. These edits are smoother than manual
replacement but usually lack learned font synthesis, leaving residual
typographic and texture inconsistencies.

\subsection{AI-Assisted Forgery}

AI-assisted forgeries introduce learned generative models for localised
document manipulation. Unlike copy-move attacks, these methods can
synthesise new visual content, reducing traditional splicing cues while
enabling higher-fidelity face replacement, portrait morphing, and
style-aware text-field editing~\cite{xie2024idnet}.

For text manipulation, ForgeNet~\cite{zhao2021deep} illustrates this
capability through modules for text editing, print-scan post-conditioning,
and halftone removal. More recent diffusion-based text editing approaches
further improve font-aware and background-consistent synthesis. At this
capability level, text-field manipulation becomes qualitatively different
from manual replacement: font matching, spacing, and local background
continuity are learned rather than approximated. Residual traces may still
remain in local noise patterns or synthesis artefacts~\cite{chen2024multi},
although high-quality reprinting and lamination can suppress some of these
cues~\cite{boned2024synthetic}.

For portrait manipulation, face morphing and face-swapping methods can
produce document portraits that retain biometric plausibility for more
than one identity, directly threatening face-verification components of
identity verification pipelines~\cite{korshunov2025fantasyid}.

\subsection{GenAI-Driven Full-Document Forgery}

GenAI-driven synthesis introduces a distinct threat model: instead of
altering a genuine document, the attacker generates a complete synthetic
identity document from template priors. Diffusion and multimodal
generation models can reconstruct document backgrounds, approximate
layout conventions, and support field-level personalisation before the
resulting image is submitted digitally or reproduced physically
~\cite{xie2024idnet}.

The threat accelerated in 2024--2025 as multimodal generative interfaces
lowered the expertise required for document-level manipulation. Tasks that
previously required specialised image-editing skill can increasingly be
specified through natural language or image-conditioned editing workflows.
This represents a capability shift: sophisticated document manipulation
becomes accessible to a broader set of attackers.

Unlike earlier classes, fully synthetic documents need not contain any
authentic substrate, issuance process, or genuine physical security
feature. This weakens forensic assumptions based on reflection analysis,
hologram inspection, substrate texture, or print-scan artefacts. Detection
therefore requires substrate-independent evidence, including structural
consistency, semantic validity, template compliance, and cross-field
reasoning. Section~\ref{sec:genai} analyses these generative AI
capabilities and limitations in detail.

\subsection{Risk--Capability--Impact Mapping}

The four forgery classes differ in capability requirements, operational
status, detection difficulty, and verification impact. Table
~\ref{tab:forgery_mapping} summarises these differences. Text-field
manipulation appears across multiple classes, illustrating how a single
attack modality can range from trivial manual editing to high-fidelity
generative synthesis depending on the available tools.

\begin{table*}[tb]
\centering
\caption{Risk--Capability--Impact Mapping of Identity Document Forgery Classes}
\label{tab:forgery_mapping}
\setlength{\tabcolsep}{5pt}
\renewcommand{\arraystretch}{1.2}
\begin{tabular}{p{2.8cm}p{3.2cm}p{2.8cm}p{3.4cm}p{1.8cm}}
\toprule
\textbf{Forgery Class} & \textbf{Capability Barrier} &
\textbf{Operational Status} & \textbf{Detection Difficulty} &
\textbf{Impact Level} \\
\midrule
Opportunistic &
Low (basic editing or reproduction) &
Common in low-effort attacks &
Low--Moderate (global classifiers, frequency analysis) &
Moderate \\
\midrule
Structure-Preserving &
Moderate (template knowledge, compositing tools) &
Common in targeted fraud &
Moderate (edge, noise, and consistency analysis) &
High \\
\midrule
AI-Assisted &
Moderate--High (specialised tools; declining barrier) &
Emerging and accelerating &
High (synthesis suppresses traditional cues) &
Very High \\
\midrule
GenAI-Driven &
Low--Moderate (prompt and image-conditioned interfaces) &
Emerging; rapidly evolving &
Very High (weak substrate and recapture assumptions) &
Critical \\
\bottomrule
\end{tabular}
\end{table*}

This mapping highlights an asymmetric transition in the threat landscape.
Opportunistic and structure-preserving forgeries remain operationally
important, but AI-assisted and GenAI-driven attacks challenge detection
methods that rely primarily on physical artefacts~\cite{chen2022domain,
mudgalgundurao2022pixel,korshunov2025fantasyid}. Future detection systems
must therefore combine visual artefact analysis with semantic consistency,
template compliance, substrate-independent liveness cues, and multi-layer
verification architectures.

\section{Detection Methods: From Heuristics to Foundation Models} 
\label{sec:detection}

\subsection{Rule-Based and Heuristic Methods}

Early identity document forgery detection, prior to deep learning, relied on handcrafted features and explicit decision rules to identify straightforward manipulation attempts. These methods exploited visually observable or structurally constrained document properties through several complementary strategies.

Security information embedding approaches used digital watermarking and steganography to verify document authenticity by comparing extracted hidden data with visible document information \cite{perry2000digital,thongkor2012digital}, though these remained vulnerable to removal and cryptographic attacks. Character-level analysis techniques detected inconsistencies through Bounding Box (BB) examination of height, orientation, and thickness variations \cite{patgar2014unsupervised}, while Scan-Edit-Print (SEP) methods identified copy-paste forgeries and font property imitations using intrinsic document features \cite{bertrand2013system}. Copy-Move Forgery Detection (CMFD) combined OCR-based analysis of font weight, size, and style with background texture examination to expose duplicated content \cite{abramova2016detecting}. Printer classification approaches distinguished forgery through printing artefacts, employing SVM-based edge sharpness analysis \cite{lampert2006printing}, DCT-based texture feature extraction \cite{schulze2009using}, and unsupervised anomaly detection without prior training \cite{gebhardt2013document}. Document layout verification methods analysed text alignment, skew, and line spacing to detect formatting violations \cite{van2013text}.

A related but distinct branch addressed document quality verification as a proxy for forgery detection. Al-Ghadi et al. \cite{al2025idtrust} introduced bandpass filtering to isolate Guilloche background degradation patterns and distinguish original from scanned IDs, eliminating the need for reference templates. While effective for print-scan discrimination, such methods remained global and did not localize manipulated regions.

While computationally efficient and interpretable, these rule-based methods exhibited limited robustness against skilled forgeries and diverse document templates. Their dependence on manually engineered features and fixed thresholds restricted scalability and generalization, motivating the subsequent shift toward learning-based detection paradigms.

\begin{table}[ht]
\centering
\caption{Summary of Rule-Based and Heuristic Detection Methods}
\label{tab:rulebased}
\resizebox{\columnwidth}{!}{%
\begin{tabular}{p{2.8cm} p{3.0cm} p{3.0cm} p{3.2cm}}
\toprule
\textbf{Method} & \textbf{Feature Type} & \textbf{Attack Targeted} & \textbf{Key Limitation} \\
\midrule
Watermarking / Steganography \cite{perry2000digital,thongkor2012digital} &
Hidden data embedding &
Authenticity verification &
Vulnerable to removal attacks \\
\midrule
Bounding Box / Character Analysis \cite{patgar2014unsupervised} &
Font height, orientation, thickness &
Text-level manipulation &
Fails on skilled font imitation \\
\midrule
Scan-Edit-Print (SEP) \cite{bertrand2013system} &
Intrinsic document features &
Copy-paste forgeries &
Limited to print-scan pipeline \\
\midrule
CMFD / OCR Analysis \cite{abramova2016detecting} &
Font weight, background texture &
Duplicated content &
Sensitive to template variation \\
\midrule
Printer Classification \cite{lampert2006printing,schulze2009using,gebhardt2013document} &
Edge sharpness, DCT texture, anomaly scores &
Print artefact detection &
Fixed thresholds, poor generalization \\
\midrule
Layout Verification \cite{van2013text} &
Text alignment, skew, line spacing &
Formatting violations &
No localisation capability \\
\midrule
Bandpass / Guilloche Filtering \cite{al2025idtrust} &
Frequency-domain degradation &
Print-scan discrimination &
Global only, no region localisation \\
\bottomrule
\end{tabular}%
}
\end{table}


\subsection{Global Deep Learning Classification}

Deep learning fundamentally reshaped the landscape of presentation attack detection for identity documents \cite{ruiz2025identity}, enabling models to autonomously learn highly discriminative representations directly from document images. Early deep learning models employed global classification techniques that operated at the image level, classifying entire documents as either bona fide or attack based on holistic features extracted from the full document image \cite{gonzalez2025forged, gonzalez2020hybrid, gonzalez2022towards}. 

A prominent trend evolved toward hybrid and multi-stage architectures that decompose the detection problem into specialised subtasks, assigning dedicated neural networks to different attack families. Gonzalez et al. \cite{gonzalez2020hybrid} pioneered a two-stage system where the first stage uses MobileNet to distinguish composite (digitally manipulated) from non-composite images, followed by a second stage where BasicNet employs Discrete Fourier Transform (DFT) preprocessing to classify documents as real, printed, or screen-displayed. This architecture was subsequently refined by incorporating Face Image Quality Assessment (FIQA) using MobileNetV2 \cite{gonzalez2022towards}, where MagFace \cite{meng2021magface} was employed to evaluate face photograph quality on the documents. This work demonstrated that system performance can be significantly improved when face quality is taken into consideration during the detection process. The approach culminated in a comprehensive system trained on 190,000 images \cite{gonzalez2025forged}, which expanded the attack taxonomy to include synthetic and PVC card attacks, while also contributing PyPAD~\footnote{\url{https://github.com/jedota/PyPAD}} — an ISO/IEC 30107-3 compliant evaluation framework \cite{ISO30107-3-2023} for standardized performance assessment.

In screen recapture detection, Chen et al. \cite{chen2025moire} proposed Moiré Spectral Augmentation (FMAG) and Masked Moiré Frequency Modeling (M$^2$FM), leveraging frequency-domain prior knowledge of display-camera distortion. By explicitly modeling moiré spectral peaks and reconstructing suppressed high-frequency components in low-quality samples, their approach significantly improved cross-domain generalization under screen-recapture scenarios. However, the framework remains image-level and does not provide spatial localisation of manipulated regions.

While global PAD-style classification achieves strong performance under controlled conditions, its primary limitation lies in the absence of spatial interpretability. Decisions are made at the image level, providing no indication of the manipulated region, which limits forensic analysis and human verification. Additionally, these models exhibit limited generalization across document types, capture devices, and unseen attack instruments \cite{gonzalez2025forged, mudgalgundurao2022pixel}, with competition benchmarks revealing EERs of 21.87\% and 11.34\% in cross-dataset scenarios \cite{tapia2024first,tapia2025second}. This led to the need for local region-specific analysis, pixel-level detection, and forensic-aware models.

\begin{table}[ht]
\centering
\caption{Summary of Global Deep Learning Classification Methods}
\label{tab:globaldl}
\resizebox{\columnwidth}{!}{%
\begin{tabular}{p{2.8cm} p{2.8cm} p{2.5cm} p{3.2cm}}
\toprule
\textbf{Method} & \textbf{Architecture} & \textbf{Attack Types} & \textbf{Key Limitation} \\
\midrule
Gonzalez et al. \cite{gonzalez2020hybrid} &
Two-stage MobileNet + BasicNet (DFT) &
Print, Screen, Composite &
Private dataset, no localisation \\
\midrule
Gonzalez et al. \cite{gonzalez2022towards} &
Two-stage MobileNetV2 + MagFace FIQA &
Print, Screen, Composite &
Face quality dependency \\
\midrule
Gonzalez and Tapia \cite{gonzalez2025forged} &
MobileNetV2, trained on 190K images &
Print, Screen, Synthetic, PVC &
Poor cross-dataset generalization \\
\midrule
Chen et al. \cite{chen2025moire} &
FMAG + M$^2$FM frequency modeling &
Screen recapture &
Image-level only, no localisation \\
\bottomrule
\end{tabular}%
}
\end{table}

\subsection{Local and Forensic-Based Detection and Localisation}

The limitations of global classification approaches motivated a shift toward methods operating at finer granularities, analysing pixel-level patterns and microscale artefacts introduced by printing, scanning, recapture, or digital injection processes. The methods reviewed in this subsection address three distinct attack surfaces: recapture and print-scan artefacts, edge and boundary inconsistencies from digital or physical text-level manipulation, and physical occlusion attacks, each requiring different detection strategies.

\subsubsection{Forensic Micro-Artifact Analysis}

A core line of research focuses on detecting unavoidable forensic traces that physical attack processes leave on document images, including moiré patterns, compression artefacts, paper textures, and colour distortions. Magee et al. \cite{magee2023investigation} investigated the Meijering filter for recapture detection, extracting textural features from grayscale images that magnify differences between screen-recaptured and paper-printed documents, demonstrating that specialised filters combined with SVM classifiers can effectively distinguish attack types without deep feature extractors.

From a deep learning perspective, Mudgalgundurao et al.~\cite{mudgalgundurao2022pixel} proposed a DenseNet-based architecture with dual supervision combining image-level binary classification with pixel-wise annotation, enabling detection of low-level moiré effects and print artefacts without explicit preprocessing. Chen et al.~\cite{chen2022domain} advanced forensic comparison through a triplet network using a ResNeXt101 backbone, learning an embedding space where genuine document patches cluster near high-quality references while recaptured patches are pushed away, preserving the forensic signature of each document type across varying capture devices and printing substrates.

Building on multimodal forensic reasoning, Chen et al.~\cite{chen2024multi} introduced MMDT, which disentangles recaptured traces into blur content and texture components via a self-supervised network, treating these as independent modalities alongside RGB and fusing them through Adaptive Multi-Modal Adapters within a ViT-B16 transformer. Li et al.~\cite{li2023two} similarly proposed a two-branch architecture processing DCT-based frequency features and RGB colour distortions in parallel, fused through multi-scale cross-attention, demonstrating improved robustness across cross-dataset and cross-quality scenarios.

Recent works extend this trajectory toward boundary-aware and edge-sensitive modeling. Bae et al. \cite{bae2025enhancing} proposed plug-and-play Edge Attention (EA) and Edge Concatenation (EC) modules that inject Sobel-derived edge maps into deep network feature maps, reinforcing sensitivity to asymmetric boundary inconsistencies introduced during physical or digital text-level and character-level manipulations. Evaluated across CNN, ResNet50, DenseNet121, ViT, and CAE-SVM backbones, these modules demonstrate consistent gains on document forgery benchmarks including DocTamper and MIDV-2020. Complementing this, IML-ViT \cite{ma2023iml}, originally proposed for general image manipulation localisation, introduced a high-resolution Vision Transformer with multi-scale supervision and morphological edge loss, and serves as a methodological reference demonstrating that transformer self-attention is particularly suited to modeling non-semantic artefact discrepancies at fine spatial scales without handcrafted noise filters.

A third and structurally distinct attack surface involves physical occlusion, where document information is concealed through overlaid physical objects rather than digital synthesis or recapture. Zhu et al. \cite{zhu2025towards} addressed physical occlusion attacks where stickers, barcodes, or overlays are used to conceal document information. Their Adversary Decomposition Network (ADDNet) disentangles document-dependent features from occlusion-specific features via adversarial classifiers, while a lightweight Document Adapter enables few-shot adaptation to unseen document types. Unlike injection or recapture attacks, physical occlusion introduces structured spatial discontinuities without synthesis artefacts, expanding the attack surface beyond digitally detectable traces. Critically, all methods discussed above, whether forensic, edge-sensitive, or occlusion-aware, rely on some form of physically introduced artefact. The following subsection addresses injection attacks, where no physical substrate is involved and artefacts are actively suppressed through generative blending, rendering these forensic approaches fundamentally insufficient.

\begin{table}[tb]
\centering
\caption{Summary of Local and Forensic-Based Detection Methods}
\label{tab:forensic}
\resizebox{\columnwidth}{!}{%
\begin{tabular}{p{2.8cm} p{2.8cm} p{2.8cm} p{2.0cm} p{2.2cm}}
\toprule
\textbf{Method} & \textbf{Architecture} & \textbf{Detection Signal} & \textbf{Localisation} & \textbf{Attack Surface} \\
\midrule
Magee et al. \cite{magee2023investigation} &
Meijering filter + SVM &
Textural frequency differences &
No &
Screen recapture \\
\midrule
Mudgalgundurao et al. \cite{mudgalgundurao2022pixel} &
DenseNet, dual supervision &
Moiré, print pixel artefacts &
Yes &
Print, Screen \\
\midrule
Chen et al. \cite{chen2022domain} &
Triplet network, ResNeXt101 &
Forensic patch similarity &
Partial &
Recapture \\
\midrule
Chen et al. \cite{chen2024multi} &
MMDT, ViT-B16, AMA adapters &
Disentangled blur and texture traces &
Partial &
Recapture \\
\midrule
Li et al. \cite{li2023two} &
Two-branch CNN, cross-attention &
DCT frequency and RGB distortion &
Partial &
Recapture \\
\midrule
Bae et al. \cite{bae2025enhancing} &
Edge Attention / Concatenation modules &
Sobel edge asymmetry &
Yes &
Text manipulation \\
\midrule
IML-ViT \cite{ma2023iml} &
High-resolution ViT, morphological loss &
Non-semantic artefact discrepancies &
Yes &
General manipulation \\
\midrule
Zhu et al. \cite{zhu2025towards} &
ADDNet, adversarial decomposition &
Occlusion discontinuities &
Yes &
Physical occlusion \\
\bottomrule
\end{tabular}%
}
\end{table}

\subsection{Specialised Digital Injection Attack Detection}

Digital injection attacks constitute a distinct threat vector because
manipulated or synthetic document images are submitted directly into KYC
verification pipelines without physical recapture
\cite{korshunov2025fantasyid}. Unlike presentation attacks, these
forgeries may contain no substrate, print-scan, or sensor-level artefacts.
Face-region replacement, portrait morphing, and text-field inpainting can
therefore suppress many cues exploited by recapture and print-detection
methods, shifting the detection problem from physical trace analysis
toward semantic, structural, and localisation-aware reasoning.

George and Marcel \cite{george2025edgedoc} proposed EdgeDoc, a hybrid CNN-Transformer architecture integrating NoisePrint-based device fingerprints with a U-Net style decoder for joint binary classification and forgery localisation. By fusing device-level forensic traces with transformer-based patch-level reasoning, EdgeDoc demonstrated competitive performance under limited training data conditions at the ICCV 2025 DeepID Challenge \cite{korshunov2025deepid}.

The DeepID challenge \cite{korshunov2025deepid} provided broader empirical evidence of the difficulty of this problem, evaluating 26 teams on both an in-domain FantasyID set and a private dataset of 20K real ID documents. Off-the-shelf methods including TruFor~\cite{guillaro2023trufor} and MMFusion~\cite{triaridis2024exploring}, while performing reasonably on text inpainting (HTER $\approx$ 24--31\%), failed near-randomly on face swapping attacks (HTER $\approx$ 37--58\%), as Gaussian blending in face-swapping tools suppresses the cut-paste artefacts these models rely on \cite{korshunov2025fantasyid}. Successful approaches instead built complex pipelines atop these baselines: the Sunlight team employed Re-MTKD, pretraining specialist teacher models on natural image forgery datasets before ID-specific fine-tuning; UAM-Biometrics adapted TruFor~\cite{guillaro2023trufor} with a SegFormer backbone and knowledge distillation, achieving the strongest cross-domain performance on the private real-world set. Critically, no submission generalised to fine-tuned text manipulations with Poisson blending on small text fields, with MCC scores near zero across all teams, highlighting that semantically targeted manipulations with deliberate artefact suppression remain an open challenge.

These findings collectively reveal a critical distinction: while presentation attacks manifest physical-domain distortions detectable through forensic trace analysis, injection attacks suppress such artefacts through generative editing and blending, demanding models capable of reasoning about subtle semantic and structural inconsistencies rather than physical forensic cues alone.

\begin{table}[tb]
\centering
\caption{Summary of Digital Injection Attack Detection Methods}
\label{tab:injection}
\resizebox{\columnwidth}{!}{%
\begin{tabular}{p{2.8cm} p{2.8cm} p{3.0cm} p{2.8cm}}
\toprule
\textbf{Method} & \textbf{Architecture} & \textbf{Detection Signal} & \textbf{Key Finding} \\
\midrule
EdgeDoc \cite{george2025edgedoc} &
CNN-Transformer, NoisePrint, U-Net decoder &
Device fingerprints, blending inconsistencies &
Competitive under limited training data \\
\midrule
TruFor~\cite{guillaro2023trufor} / MMFusion~\cite{triaridis2024exploring} &
SegFormer, multi-stream fusion &
Noise traces, semantic mismatch &
Fails on face swap, partial on text inpainting \\
\midrule
Sunlight (Re-MTKD) \cite{korshunov2025deepid} &
Teacher-student distillation &
Cross-domain forensic features &
Strong in-domain, weak on Poisson-blended text \\
\midrule
UAM-Biometrics \cite{korshunov2025deepid} &
TruFor + SegFormer + knowledge distillation &
Structural and semantic inconsistencies &
Best cross-domain on real-world private set \\
\bottomrule
\end{tabular}%
}
\end{table}

\subsection{Foundation Models, Vision-Language Models, and Few-Shot Learning for ID Card Forgery Detection}

\begin{table}[ht]
\centering
\caption{Summary of Foundation Model, VLM, and Few-Shot Learning Methods for ID Card PAD}
\label{tab:fmvlmfsl}
\resizebox{\columnwidth}{!}{%
\begin{tabular}{p{2.3cm} p{2.4cm} p{2.7cm} p{2.9cm} p{2.5cm}}
\toprule
\textbf{Method} & \textbf{Model} & \textbf{Protocol} & \textbf{Key Result} & \textbf{Key Gap} \\
\midrule
Tapia and Busch \cite{tapia2025can} &
DINOv2, CLIP &
Zero-shot and fine-tuning &
EER 4.33\% zero-shot, 8.25\% after score fusion &
No forensic pre-training \\
\midrule
Munoz-Haro et al. \cite{munoz2025fakeidet} &
DINOv2, patch-based &
Fine-tuning on anonymised patches &
0\% EER on unseen DLC-2021 &
Spanish IDs only, no digital attacks \\
\midrule
Li et al. \cite{li2026layout} &
DINOv2, SimMIM, ArcFace &
Self-supervised + metric learning &
276 fraud cases surfaced, 222 missed by incumbents &
Trained on US IDs only \\
\midrule
Sanchez et al. \cite{sanchez2024few} &
Prototypical Networks, EfficientNetV2-B0 &
FSL, episodic training &
Competitive EER with 5 identities per country &
Synthetic data only \\
\midrule
Rocamora et al. \cite{rocamora2026simulation} &
Prototypical Networks, EfficientNetV2-B0 &
FSL on real ID cards &
EER 3.10\%, BPCER$_{20}$ 2.80\% with 50 identities &
Screen and print attacks only \\
\midrule
Vidit et al. \cite{vidit2025detecting} &
GPT-4o, FakeShield, SIDA &
Zero-shot VLM benchmarking &
GPT-4o best, fine-tuned VLMs fail to generalise &
No localisation output \\
\midrule
Zeng et al. \cite{zeng2026vision} &
SmolVLM2 (500M, 2.2B), LoRA &
Generative and discriminative fine-tuning &
EER 0.93\% on Chile, 5.99\% on Mexico &
Unstable on synthetic datasets \\
\bottomrule
\end{tabular}%
}
\end{table}

The persistent limitations of task-specific CNN architectures in cross-domain generalisation have motivated a paradigm shift in ID card PAD. Recent research increasingly focuses on Foundation Models (FMs), Vision--Language Models (VLMs), and Few-Shot Learning (FSL). Unlike conventional deep learning models trained from scratch on scarce domain-specific data, these approaches provide principled pathways toward improved generalisation. They also reduce the need for large collections of sensitive identity documents.

Tapia and Busch~\cite{tapia2025can} benchmarked DINOv2~\cite{oquab2023dinov2} and CLIP~\cite{radford2021learning} for ID card PAD under zero-shot and fine-tuning protocols, finding that DINOv2 consistently outperformed both CLIP and conventional vision transformers, achieving EERs as low as 4.33\% in zero-shot scenarios. A key finding was that the quality of bona fide images is more critical to generalisation than the volume of attack samples. A score-level fusion of DINOv2 with a traditional CNN further reduced EER on IDNet from 27.86\% to 8.25\%, exploiting the complementarity between local FM features and global CNN features.

Complementing this, Munoz-Haro et al.~\cite{munoz2025fakeidet} introduced FakeIDet, a privacy-preserving patch-based framework applying DINOv2 to anonymised patches from real Spanish identity documents, achieving 0\% EER at the ID level on the unseen DLC-2021 dataset while remaining fully GDPR-compliant.

Extending the FM line to open-set fraud discovery, Li et al.~\cite{li2026layout} adapted DINOv2 to the identity document domain via context-aware SimMIM fine-tuning and supervised metric learning with a composite loss combining ArcFace, supervised contrastive, and center loss objectives. Trained exclusively on US ID cards, the model achieves 99.83\% layout classification accuracy on unseen Canadian layouts. On a real-world dataset of 20,448 Canadian IDs, embedding-space clustering surfaced 276 adaptive physical fraud cases, of which 222 were missed by existing detectors. This work shifts the framing from closed-set binary classification toward open-set fraud discovery, where novel fabrication pipelines can be identified without prior labels.

On the data-efficient adaptation front, Sanchez et al.~\cite{sanchez2024few} proposed an FSL approach using Prototypical Networks with an EfficientNetV2-B0 backbone to extend PAD capabilities to previously unseen ID card countries. By learning class prototypes from a small support set, the method achieved competitive EERs with as few as five unique identities per new country across Spain, Chile, Argentina, and Costa Rica, demonstrating that meta-learning offers a practical path for rapid deployment in multi-country verification systems without requiring full retraining.

Most recently, Rocamora et al.~\cite{rocamora2026simulation} extended Prototypical Network-based FSL from synthetic documents to real government-issued ID cards, validating it against screen display and colour print attacks across four Hispanic American countries: Chile, Honduras, Nicaragua, and El Salvador. Using an EfficientNetV2-B0 backbone, the method demonstrated that just 50 unique identities from a new country suffice to achieve competitive performance, reporting an average EER of 3.10\% and BPCER$_{20}$ of 2.80\%, closely matching fully supervised baselines. Print-based attacks proved easier to detect than screen-based ones, attributed to the distinctiveness of physical artefacts such as ink diffusion and paper texture at the prototype level. This work directly addresses the simulation-to-production gap in prior FSL approaches, confirming reliability under real operational constraints.

Vidit et al.~\cite{vidit2025detecting} benchmarked VLMs for text manipulation detection on the FantasyID injection attack dataset. GPT-4o substantially outperformed open-source alternatives, leveraging semantic reasoning to detect font inconsistencies and logical contradictions beyond pixel-level forensics. Notably, task-specific fine-tuned VLMs such as FakeShield and SIDA failed to generalise to text manipulation in identity documents, and input image resolution emerged as a critical factor for detecting artefacts in constrained text fields.

Zeng et al.~\cite{zeng2026vision} (2026) proposed a compact multimodal framework for cross-domain PAD on ID cards, adapting SmolVLM2 into both generative and discriminative classification structures. The generative structure, fine-tuned via LoRA on genuine Chilean and Mexican ID cards, achieved an EER of 0.93\% on Chile and 5.99\% on Mexico, outperforming DenseNet and unimodal baselines in cross-country settings. A key finding is that zero-shot multimodal models fail entirely on this task, with EERs exceeding 45\%, confirming that general-purpose pretraining alone is insufficient for PAD without task-specific supervision. Performance on synthetic passport datasets from Poland, Portugal, and Spain was consistently unstable across all models, reinforcing the synthetic utility gap identified in the broader literature.

Collectively, these works establish FMs, VLMs, and FSL as promising but
still incomplete directions for identity document forgery detection.
Foundation models improve transfer under limited labelled data, few-shot
methods support rapid adaptation to new countries and templates, and VLMs
introduce semantic reasoning beyond pixel-level forensics. However, no
current approach resolves all operational requirements: forensic
pre-training for document artefacts remains limited, localisation from
VLMs is unreliable, zero-shot multimodal inference remains unstable across
datasets, and non-Latin scripts are still underexplored. Existing FSL
studies also focus mainly on presentation attacks, leaving few-shot
adaptation to digital injection attacks largely open. These limitations
connect directly to the dataset and benchmark gaps analysed in
Section~\ref{sec:benchmarks}.


\begin{table*}[ht]
\centering
\caption{High-Level Overview of Identity Document Forgery Detection Paradigms}
\label{tab:combined_overview}
\begin{tabular}{p{6.0cm}  p{7.0cm} p{1.5cm}}
\toprule
\textbf{Detection Paradigm} &
\textbf{Key Strength} &
\textbf{Localisation} \\
\midrule
Rule-Based \& Heuristic Methods
\cite{perry2000digital,abramova2016detecting,lampert2006printing,al2025idtrust} &
Interpretable and computationally efficient &
No \\
\midrule
Global Deep Learning Classification
\cite{gonzalez2025forged,gonzalez2020hybrid,gonzalez2022towards,chen2025moire} &
Strong closed-set PAD performance &
No \\
\midrule
Forensic Micro-Artifact Analysis
\cite{magee2023investigation,chen2024multi,li2023two,chen2022domain} &
Strong forensic sensitivity across recapture attacks &
Partial \\
\midrule
Localisation-Oriented Detection
\cite{mudgalgundurao2022pixel,bae2025enhancing,ma2023iml} &
Spatial localisation and interpretability &
Yes \\
\midrule
Physical Occlusion Attack Detection
\cite{zhu2025towards} &
Adaptation to unseen physical obstruction attacks &
Yes \\
\midrule
Digital Injection Attack Detection
\cite{george2025edgedoc,korshunov2025deepid} &
Detection without physical forensic cues &
Yes \\
\midrule
Foundation Models \& VLMs
\cite{tapia2025can,munoz2025fakeidet,vidit2025detecting,zeng2026vision} &
Strong generalisation under limited training data &
Limited \\
\midrule
Open-Set Fraud Discovery
\cite{li2026layout} &
Surfaces unseen fraud families without prior labels &
No \\
\midrule
Few-Shot \& Data-Efficient Learning
\cite{sanchez2024few,rocamora2026simulation} &
Rapid deployment to unseen countries and templates &
No \\
\bottomrule
\end{tabular}
\end{table*}

\section{Datasets, Evaluation Protocols, and the Reality Gap}
\label{sec:benchmarks}

\subsection{Synthetic Data for Training and Testing}

The primary driver behind synthetic data generation is the acute scarcity of publicly available ID card images. Privacy regulations such as the General Data Protection Regulation (GDPR)~\cite{gdpr2016} restrict the collection and dissemination of genuine identity documents, limiting the availability of large-scale datasets for research. Furthermore, the inherent class imbalance between abundant attack samples and scarce bona fide samples can degrade the performance and generalisation ability of machine learning models~\cite{wang2021review}. Synthetic data offers a way to sidestep these constraints by automatically producing both bona fide samples and presentation attack instruments (PAIs), i.e., artefacts used to spoof biometric or document verification systems, at scale without the logistical and ethical difficulties of real-world data collection.

Several generative approaches have been applied in the ID card domain. Classical texture transfer and procedural algorithms overlay device-specific noise patterns onto clean templates to simulate print and screen artefacts. Among deep generative models, CycleGAN~\cite{zhu2017unpaired} has been the most widely adopted, performing unpaired image-to-image translation to transform bona fide document images into realistic attack samples without requiring paired training data. Paired conditional GANs such as pix2pix~\cite{isola2017image} offer finer structural control over generated attacks, while StyleGAN2-ADA~\cite{karras2020training} has been applied for full-document bona fide synthesis in small dataset regimes. More recently, few-shot synthetic pipelines incorporating diffusion models have been introduced for large-scale document template generation~\cite{xie2024idnet}.

Benalcazar et al.~\cite{benalcazar2023synthetic} conducted the most comprehensive comparison of these methods, testing procedural texture addition, texture transfer, CycleGAN, and StyleGAN2-ADA, finding that texture transfer and CycleGAN produced the most forensically useful training samples. Markham et al.~\cite{markham2024open} extended this on open-source datasets, confirming CycleGAN's effectiveness for open-set scenarios. Chen et al.~\cite{chen2024distortion} took an adversarial approach, embedding a CycleGAN-based distortion generator directly into the training loop, forcing the PAD model to continuously adapt to new synthetic attacks and thereby learn more generalisable features.

\subsection{Public Datasets and Their Limitations}

Public datasets for ID card forgery detection remain limited in scale, diversity, and attack coverage, 
largely due to privacy regulations that prevent the use of real identity documents in research. 
As summarised in Table~\ref{tab:dataset_summary}, the available benchmarks vary significantly in 
their design objectives and forensic realism.

\begin{table*}[ht]
\centering
\caption{Public ID-Card Forgery Datasets (2019--2025)}
\label{tab:dataset_summary}
\begingroup
\footnotesize
\setlength{\arrayrulewidth}{0.4pt}
\setlength{\tabcolsep}{4pt}
\renewcommand{\arraystretch}{1.2}

\begin{tabular}{|p{1.8cm}|p{1.9cm}|p{2.6cm}|p{1.9cm}|p{1.1cm}|p{3.95cm}|}
\hline
\textbf{Dataset (Year)} &
\textbf{Size \& Modality} &
\textbf{Attack Taxonomy} &
\textbf{Bona Fide Source} &
\textbf{Real / Synth.} &
\textbf{Key Limitations} \\
\hline

MIDV Family (2019--2023) 
\cite{arlazarov2019midv, bulatovich2022midv, bulatov2020midv, chernyshova2021midv, koliaskina2023midv} &
$\sim$76K images \& clips (Image + Video) &
PA-Phy (MIDV-Holo: Print, Pseudo-hologram); no PA-Dig/PA-Gen/PA-Inj &
Laminated paper mock-ups &
Synthetic &
No real IDs; lacks digital composite and injection attacks; custom holograms (non-government); no pixel-level forgery ground truth; limited forensic realism \\
\hline

DLC2021 (2021) \cite{polevoy2022document} &
1,424 clips (Video) &
PA-Phy (Print, Screen) &
Laminated MIDV-2020 mock-ups &
Synthetic &
Low-complexity physical attacks only; no PA-Dig/PA-Gen; lacks genuine OVDs/holograms; no injection modelling \\
\hline

KID34K (2023) \cite{park2023kid34k} &
34,662 images (Image) &
PA-Phy (Print, Screen) &
Custom plastic cards (synthetic identities) &
Synthetic &
Country-specific (South Korea); no digital composite or injection; non-ICAO compliant; limited document diversity \\
\hline

SIDTD (2024)~\cite{boned2024synthetic} &
$\sim$75K images \& clips (Image + Video) &
PA-Dig (Crop--Replace, Inpainting) with physical re-capture variants &
MIDV-derived digital templates &
Synthetic &
Simple digital edits only; no PA-Gen/PA-Inj; limited physical security simulation; no semantic GenAI manipulation \\
\hline

IDNet (2024) \cite{xie2024idnet} &
837,060 images (Image) &
PA-Dig (Composite, Morphing, Text Rewrite, Mixed); no PA-Phy/PA-Inj &
AI-generated synthetic templates &
Synthetic &
Fully synthetic; lacks physical capture artefacts; forensic gap to real IDs; no physical presentation attacks \\
\hline

Syn-IDPASS (2025) \cite{tapia2025synid} &
9,000 images (Image) &
PA-Phy (Print, Screen) &
ICAO-9303 compliant synthetic passports &
Synthetic &
Passports only; narrow geography (3 countries); no PA-Dig/PA-Gen; lacks real physical security elements \\
\hline

RSCID (2022) \cite{chen2022domain} &
1,104 images (Image) &
PA-Phy (Print--Scan, Screen, Substrate) &
Acrylic plastic student ID cards &
Synthetic &
Student IDs only; small scale; no digital composite or GenAI attacks; non-ICAO compliant; limited diversity \\
\hline

FantasyID (2025) \cite{korshunov2025fantasyid} &
1,086 images (Image) &
PA-Gen / PA-Inj (Face Swap, Diffusion-based Text Editing) &
Physically printed synthetic plastic cards &
Synthetic &
Injection-focused only; excludes physical replay attacks; non-ICAO compliant; limited attack diversity \\
\hline

FakeIDet-db (2025) \cite{munoz2025fakeidet} &
48,400 patches (Patch-based Image) &
PA-Phy (Print, Screen) &
Anonymized patches from real Spanish IDs &
\textbf{Real} (patch-level) &
Patch-only; no global layout analysis; no PA-Dig/PA-Gen; geographically limited (Spain); small subject count ($n$=30) \\
\hline

\end{tabular}
\endgroup

\smallskip
\noindent\textbf{Note:} PA-Phy = Physical Presentation Attack; 
PA-Dig = Digital Composite Attack; 
PA-Gen = Generative AI-based Manipulation; 
PA-Inj = Injection Attack.
Only FakeIDet-db contains data derived from real identity documents; all others rely on mock-ups or fully synthetic templates.
\end{table*}

The \textbf{MIDV family}~\cite{arlazarov2019midv, bulatovich2022midv, bulatov2020midv, chernyshova2021midv, koliaskina2023midv} represents the most sustained effort in public ID document benchmarking, evolving from basic video-based analysis to multi-script OCR and hologram 
detection, though all documents remain laminated paper mock-ups without real security features. 
\textbf{DLC2021}~\cite{polevoy2022document} builds directly on MIDV-2020 mock-ups, focusing on lamination-based 
liveness cues against simple print and screen attacks, but excludes any digital or injection attack 
scenario. \textbf{KID34K}~\cite{park2023kid34k} addresses the South Korean market with physically manufactured 
plastic cards, yet remains geographically narrow and restricted to print and screen attacks only. 
\textbf{SIDTD}~\cite{boned2024synthetic} extends MIDV-2020 with basic digital composite attacks via Crop \& Replace 
and inpainting, but the forgeries are simple pixel-level edits lacking any semantic or GenAI-driven 
manipulation. \textbf{IDNet}~\cite{xie2024idnet} offers the largest scale with over 837,000 images across 20 
document types and six attack variants, though being fully synthetic it lacks the physical capture 
artefacts of real-world scenarios. \textbf{Syn-IDPASS}~\cite{tapia2025synid} provides ICAO-9303 compliant 
synthetic passports for three European nationalities, but covers only print and screen attacks with 
no digital edits. \textbf{RSCID}~\cite{chen2022domain} explicitly targets cross-domain generalisation in 
recapture detection using physically manufactured acrylic student cards, but is limited to 
recapture attacks and does not represent government-issued documents. \textbf{FantasyID}~\cite{korshunov2025fantasyid} 
is one of the few datasets to model GenAI-driven injection attacks — face swapping and text 
inpainting — on physically printed cards, though it entirely excludes physical presentation attacks 
such as print and screen recaptures. Finally, \textbf{FakeIDet-db}~\cite{munoz2025fakeidet} is the only public 
dataset derived from real identity documents, using pseudo-anonymised patch extraction for 
privacy preservation, but its patch-only format precludes any analysis of global document layout 
or context.

Across all these resources, several critical gaps persist. No dataset simultaneously covers physical presentation attacks, digital composite forgeries, and GenAI-driven full-document synthesis. Localisation ground truth — pixel- or region-level annotations marking forged areas — is absent in virtually all cases. Most bona fide samples are mock-ups or purely synthetic templates that lack the optical and physical security features of genuine circulating documents, limiting the forensic relevance of models trained on them.

\subsection{Evaluation Protocols and Performance Metrics}

The standardised evaluation of PAD systems for identity documents is anchored in the ISO/IEC 30107-3 framework \cite{ISO30107-3-2023}, which defines the two primary error metrics used across the literature: the Attack Presentation Classification Error Rate (APCER) and the Bona Fide Presentation Classification Error Rate (BPCER). APCER measures the proportion of attack presentations incorrectly classified as bona fide, capturing the system's security vulnerability, while BPCER quantifies how often genuine presentations are wrongly rejected, reflecting user inconvenience. The Equal Error Rate (EER) — the operating point where APCER and BPCER are equal — serves as a single-value summary of the system's overall discriminative capacity and remains the most widely reported metric in the reviewed literature.
In practice, systems are evaluated at specific security-oriented operating points. Metrics such as BPCER\textsubscript{10}, BPCER\textsubscript{20}, and BPCER\textsubscript{100} report the false rejection rate at APCER thresholds of 10\%, 5\%, and 1\%, respectively. The IJCB PAD-ID-card competitions \cite{tapia2024first} formalised a composite AVRank score weighting these metrics to emphasise worst-case operational performance:
\begin{equation}
\mathrm{AVRank} = \mathrm{BPCER}_{10} \times 0.2 + \mathrm{BPCER}_{20} \times 0.3 + \mathrm{BPCER}_{100} \times 0.5
\end{equation}
This weighted formulation prioritises high-security scenarios where even low attack acceptance rates are critical \cite{tapia2025second}. Standardised tooling such as PyPAD \cite{gonzalez2025forged} facilitates reproducible reporting under this framework.

A parallel but distinct evaluation paradigm has emerged for injection and digital manipulation attacks. The DeepID Challenge at ICCV 2025 introduced F1-score as the primary metric for both binary detection and pixel-level localisation tracks, computed as a weighted aggregate over an in-domain synthetic set and an out-of-domain private set of 20K real ID documents, deliberately prioritising real-world generalisation over in-domain performance~\cite{korshunov2025deepid}. For the localisation track, the Matthews Correlation Coefficient (MCC) was additionally computed per manipulated image to account for the severe pixel-level class imbalance inherent to localisation tasks, where manipulated pixels represent only a small fraction of the total image.

For evaluating synthetic training data quality, the field commonly relies on the Fréchet Inception Distance (FID) \cite{heusel2017gans}, the Learned Perceptual Image Patch Similarity (LPIPS) metric \cite{zhang2018unreasonable}, and VGG-based perceptual loss \cite{johnson2016perceptual}. However, a critical disconnect exists between perceptual fidelity and forensic utility: visually plausible synthetic samples do not necessarily serve as effective training signal \cite{markham2024open}, underscoring the absence of a dedicated metric that jointly captures both dimensions.

Collectively, these protocols reveal a fundamental fragmentation in how the field evaluates itself: APCER/BPCER/EER dominate physical PAD benchmarks, while F1-score and MCC are better suited to digital manipulation and localisation tasks. No unified evaluation framework currently bridges both threat vectors, nor does any standardised metric capture localisation quality for physical PAD — a gap that limits both comparability and auditability across the literature.

\subsection{The Reality Gap and Implications for Benchmark Design}

A persistent mismatch exists between the controlled conditions under which current PAD models are evaluated and the complexity of real-world deployment. The IJCB PAD-ID-card competitions~\cite{tapia2024first, tapia2025second} made this gap empirically visible: models trained on private, large-scale industry datasets significantly outperformed those trained on public benchmarks, confirming that existing open datasets — composed predominantly of mock-ups or purely synthetic templates — fail to capture the forensic complexity of genuine circulating documents. Varied capture devices, unseen document templates, cross-country generalisation, and compression artefacts introduced during mobile onboarding all represent conditions that no current public dataset adequately models.

This gap has direct consequences for benchmark design. When bona fide samples are synthetic templates rather than real captured documents, classifiers risk learning template-specific features rather than genuine liveness cues. Similarly, when attack coverage is restricted to print and screen recaptures — as in the majority of available datasets — models are never exposed to digital composite forgeries, injection attacks, or GenAI-driven manipulations during training, leaving them blind to the most rapidly evolving threat vectors. The absence of pixel-level localisation annotations across all public datasets further means that no benchmark currently supports the evaluation of region-level detection, which is essential for both explainability and legal accountability. Collectively, these design choices create a self-reinforcing cycle: methods are optimised for available benchmarks, and those benchmarks do not reflect the attacks that matter most in practice.


\subsection{Zero-Shot Benchmarking of Public Models on Unseen Synthesised ID Cards}
\label{sec:zeroshot}

To provide an empirical reference point for the Reality Gap, we evaluate publicly available detectors and multimodal models under zero-shot conditions on a newly constructed evaluation set. All models were tested on document templates not seen during training, making the experiment a direct stress test of cross-domain generalisation.

\subsubsection{Dataset}
\begin{figure}[ht]
    \centering
    \begin{subfigure}[b]{0.24\linewidth}
        \includegraphics[width=\linewidth]{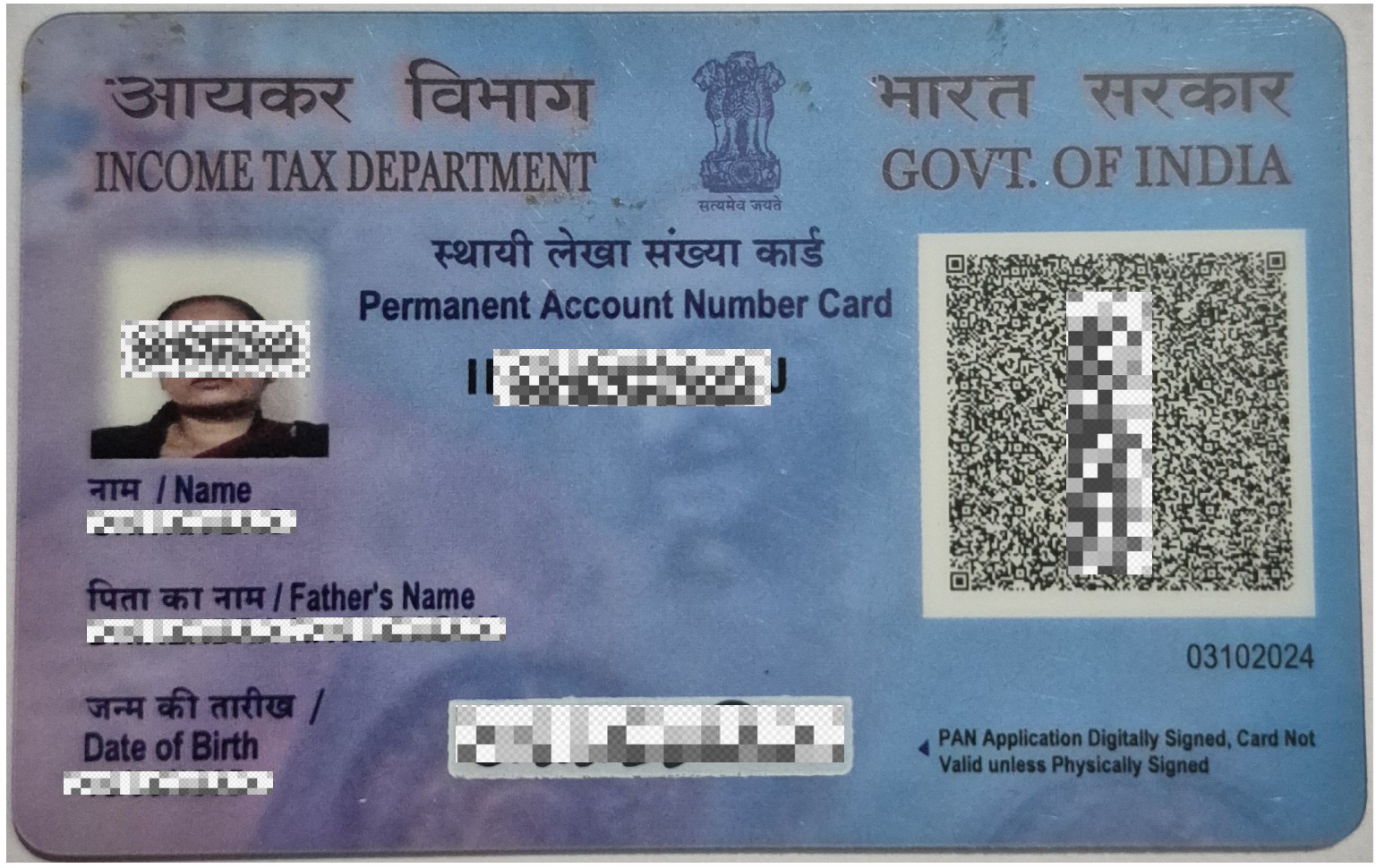}
        \caption{}
        \label{fig:sample_a}
    \end{subfigure}
    \hfill
    \begin{subfigure}[b]{0.24\linewidth}
        \includegraphics[width=\linewidth]{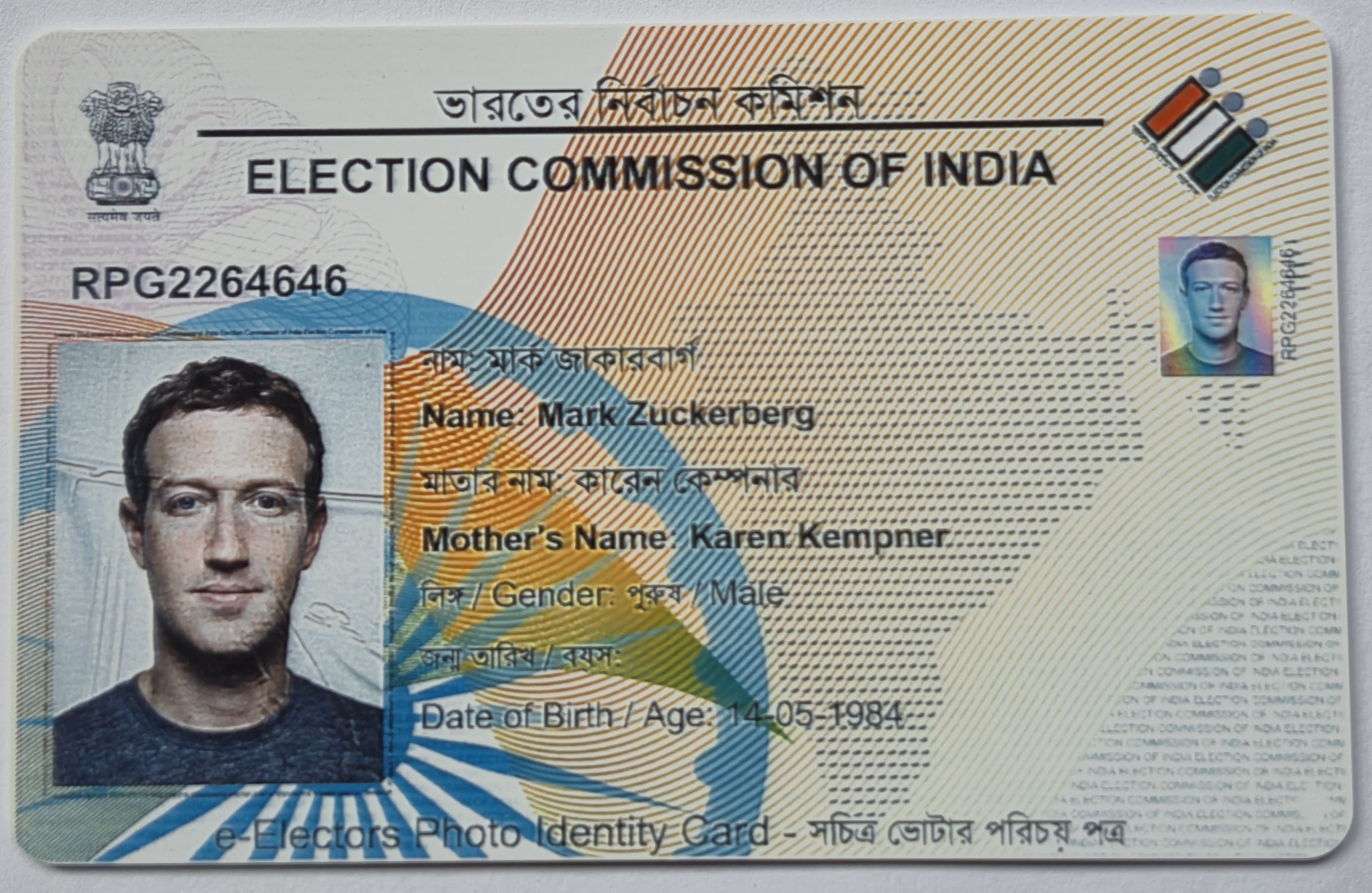}
        \caption{}
        \label{fig:sample_c}
    \end{subfigure}
    \hfill
    \begin{subfigure}[b]{0.24\linewidth}
        \includegraphics[width=\linewidth]{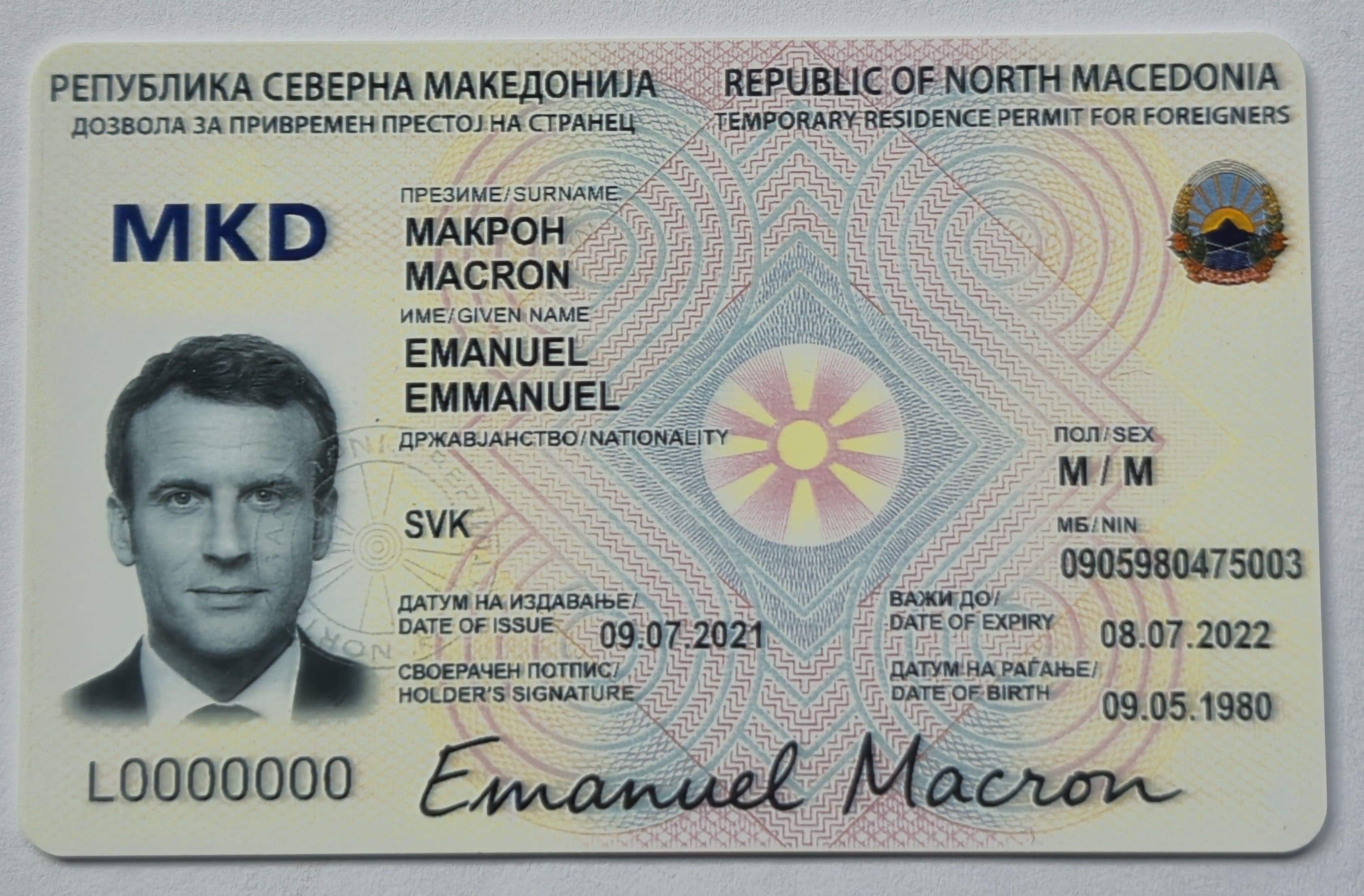}
        \caption{}
        \label{fig:sample_d}
    \end{subfigure}
    \hfill
    \begin{subfigure}[b]{0.24\linewidth}
        \includegraphics[width=\linewidth]{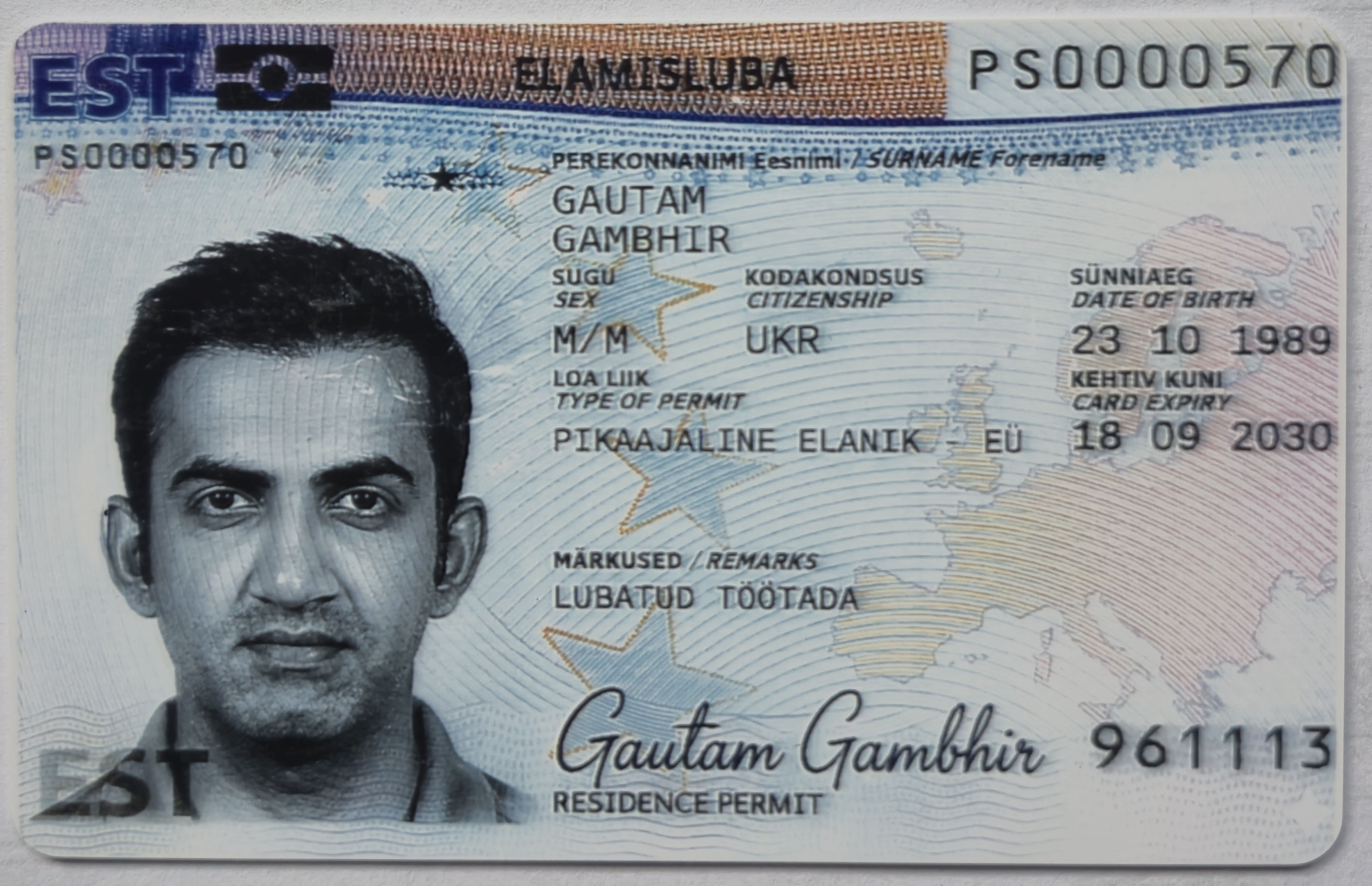}
        \caption{}
        \label{fig:sample_f}
    \end{subfigure}
    \caption{Representative samples from the dataset. (a) A genuine bona fide ID card. (b)--(d) Synthesised attack samples printed on PVC cards.}
    \label{fig:dataset_samples}
\end{figure}

The dataset contains 696 images: 560 attack images derived from 140 unique synthesised ID cards captured across four devices under similar lighting conditions, and 136 bona fide images of genuine identity documents captured under matching conditions. The bona fide images were used only for evaluation and were handled under privacy-preserving storage and access procedures. Attack templates were derived from public reference layouts, including EU PRADO entries\footnote{\url{https://www.consilium.europa.eu/prado/en/prado-latest-authentic.html}} and Indian identity document formats. Template structure was reconstructed from these references; text fields and facial photographs were replaced using generative editing, and the resulting documents were printed on PVC cards to replicate physical presentation attack conditions, following a simulation-to-production pipeline consistent with prior work ~\cite{rocamora2026simulation}. A subset of bona fide samples shares document templates with the attack set, allowing evaluation against template-matched forgeries rather than relying only on template novelty. Figure~\ref{fig:dataset_samples} shows representative samples.

\subsubsection{Evaluated Models}
Nine publicly available models are evaluated. DINOv2 ViT-L/14~\cite{oquab2023dinov2}, the Two-Branch Architecture~\cite{li2023two}, and the AS-DPAD adversarial surrogate model~\cite{chen2023distortion} were fine-tuned on subsets of DLC2021~\cite{polevoy2022document} and KID34K~\cite{park2023kid34k}. EdgeDoc~\cite{george2025edgedoc} was trained on FantasyID~\cite{korshunov2025fantasyid}, following the protocol established at the DeepID challenge~\cite{korshunov2025deepid}. The Sunlight submission was used with its publicly released weights from the same challenge. Gemini 2.5 Pro, Gemini 2.5 Flash, GPT-5, and GPT-5 Mini (tests performed on 1 June 2026) were queried in zero-shot mode without any fine-tuning.

\subsubsection{Benchmark Protocol}
All models were evaluated without exposure to the evaluation images or attack templates. For models producing continuous scores, thresholds were swept to compute APCER, BPCER, EER, and BPCER at fixed APCER operating points. For VLMs, a fixed prompt requested a binary authenticity decision with a confidence score; responses were parsed into the same score space before threshold sweeping. Metrics are reported at image level. Because the attack set is derived from 140 unique synthetic cards captured across four devices, multiple captures from the same card are not statistically independent; the results should therefore be interpreted as a zero-shot stress test of cross-template and cross-capture generalisation rather than as deployment certification. Prompts, parsing rules, and evaluation scripts are retained to support reproducibility.

\subsubsection{Results and Discussion}

\begin{table*}[tb]
\centering
\caption{Zero-shot evaluation of public models on unseen synthesised ID
cards, including bona fide presentations. APCER and BPCER are reported at
the default threshold of 0.5, while EER is reported at the equal-error
threshold. The final two columns report the mean $\pm$ standard deviation
of the predicted class probability for bona fide and attack samples.}
\label{tab:zeroshot_bench}
\resizebox{\textwidth}{!}{%
\begin{tabular}{lcccccccc}
\toprule
\textbf{Method}
  & \textbf{APCER} & \textbf{BPCER} & \textbf{EER}
  & \textbf{BPCER@1\%} & \textbf{BPCER@5\%} & \textbf{BPCER@10\%}
  & $\boldsymbol{P(\mathrm{bona\ fide})}$ & $\boldsymbol{P(\mathrm{attack})}$ \\
  & (\%) & (\%) & (\%) & (\%) & (\%) & (\%)
  & Mean $\pm$ SD & Mean $\pm$ SD \\
\midrule
GPT-5
  & 37.14 & 5.88  & 13.77
  & 78.24 & 32.35 & 26.47
  & $0.814 \pm 0.160$ & $0.574 \pm 0.284$ \\
GPT-5 Mini$^{\,\dagger}$
  & 37.14 & 0.00  & 14.71
  & 93.53 & 58.82 & 51.47
  & $0.833 \pm 0.034$ & $0.574 \pm 0.284$ \\
Gemini~2.5 Flash~\cite{comanici2025gemini}$^{\,\ddagger}$
  & 24.29 & 17.65 & 17.65
  & 73.04 & 59.31 & 42.16
  & $0.802 \pm 0.340$ & $0.700 \pm 0.368$ \\
Gemini~2.5 Pro~\cite{comanici2025gemini}
  & 30.71 & 2.94  & 28.48
  & 100.00 & 100.00 & 96.73
  & $0.932 \pm 0.156$ & $0.696 \pm 0.450$ \\
DINOv2 ViT-L/14~\cite{oquab2023dinov2}
  & 27.14 & 38.24 & 29.41
  & 97.06 & 85.29 & 73.53
  & $0.584 \pm 0.244$ & $0.641 \pm 0.256$ \\
Chen et al.~\cite{chen2023distortion}
  & 27.86 & 32.35 & 32.32
  & 79.41 & 64.71 & 49.26
  & $0.615 \pm 0.217$ & $0.619 \pm 0.205$ \\
EdgeDoc~\cite{george2025edgedoc}
  & 30.71 & 76.47 & 47.06
  & 97.79 & 97.06 & 97.06
  & $0.335 \pm 0.300$ & $0.635 \pm 0.379$ \\
Sunlight Code~\cite{korshunov2025deepid}
  & 25.00 & 61.76 & 50.71
  & 94.12 & 94.12 & 91.18
  & $0.342 \pm 0.380$ & $0.681 \pm 0.393$ \\
Li et al.\ (Two-Branch)~\cite{li2023two}
  & 32.14 & 82.35 & 52.14
  & 100.00 & 97.06 & 91.18
  & $0.277 \pm 0.328$ & $0.620 \pm 0.388$ \\
\bottomrule
\end{tabular}%
}

\smallskip
\footnotesize
\noindent
For VLMs, predicted probabilities correspond to the model's stated
confidence in its verbal decision; for CNN- and embedding-based models,
they correspond to the softmax output of the final classification layer.
BPCER@$k$\% denotes BPCER at a fixed APCER of $k$\%, equivalent to
BPCER$_{100/k}$ under ISO~30107-3 notation~\cite{ISO30107-3-2023}.
$^{\dagger}$BPCER is 0.00\% on $n=136$ bona fide samples; the
Clopper--Pearson exact 95\% confidence interval is
$[0.00\%,\,2.68\%]$.
$^{\ddagger}$BPCER equals EER, indicating that the default threshold
coincides with the equal-error operating point.
\end{table*}

Table~\ref{tab:zeroshot_bench} reports APCER, BPCER, EER, and BPCER at
fixed APCER operating points of 1\%, 5\%, and 10\%, corresponding to
BPCER$_{100}$, BPCER$_{20}$, and BPCER$_{10}$. The results show that none
of the nine evaluated models is reliable enough to act as a standalone
detector under security-oriented operating conditions. Every method
exceeds 70\% BPCER at APCER=1\%, and several remain above 90\% even at
APCER=10\%. This means that when the threshold is tightened to reject
attacks more aggressively, the models reject a large fraction of genuine
documents as well.

The results also show why aggregate metrics can be misleading. GPT-5 Mini
achieves BPCER=0.00\% at the default threshold, but its EER of 14.71\% and
BPCER$_{100}$ of 93.53\% indicate that this behaviour reflects a strong
bias toward accepting bona fide samples rather than robust discrimination.
Similarly, Gemini~2.5 Pro has a low default-threshold BPCER of 2.94\%,
but reaches BPCER$_{100}$=BPCER$_{20}$=100.00\% at stricter operating
points. These cases illustrate why EER and default-threshold results
should not be interpreted alone when evaluating security-critical PAD
systems.

The score distributions provide further insight into the failure modes.
The VLMs assign high average bona fide confidence to genuine samples, but
also give relatively high and variable confidence to attack samples,
leaving the two classes insufficiently separated. Competition- and
recapture-specific models, including EdgeDoc, Sunlight, and the
Two-Branch architecture, show the opposite tendency: they assign low
average bona fide probability even to genuine documents, which drives high
BPCER. The embedding-based models show limited separation between classes;
for example, DINOv2 produces mean scores of $0.584$ for bona fide samples
and $0.641$ for attacks, while Chen et al.\ produce $0.615$ and $0.619$,
respectively. These patterns suggest that reliable deployment would
require explicit calibration of bona fide and attack score distributions,
rather than direct thresholding of generic model outputs.

Overall, the benchmark reinforces the Reality Gap identified in
Section~\ref{sec:benchmarks}. Current forensic, embedding-based,
challenge-specific, and multimodal models provide partial discrimination
in aggregate, but they do not yet support robust standalone detection of
GenAI-assisted identity document attacks under operationally meaningful
thresholds. The failure is therefore not attributable to a single model
family; it reflects a broader mismatch between available detectors and the
synthesised attack type evaluated here.


\section{Generative AI in ID Card Forgery: Capabilities, Limitations, and Forensic Implications}
\label{sec:genai}

Accessible generative AI has introduced a qualitative shift in the threat landscape for identity document verification. Capabilities that previously required specialised editing skill, document-template knowledge, and
controlled reproduction workflows are increasingly available through image-conditioned and multimodal generation systems. The attack spectrum has therefore expanded from physical presentation attacks to targeted digital manipulation and, in some cases, full-document synthesis. This section analyses these capabilities from a defensive perspective, focusing
on the forensic artefacts and structural limitations that remain exploitable for detection.

\subsection{AI-Assisted Forgery: Face and Text Manipulation}

\begin{figure}[tb]
\centering
\begin{subfigure}[t]{0.29\linewidth}
    \centering
    \includegraphics[width=\linewidth]{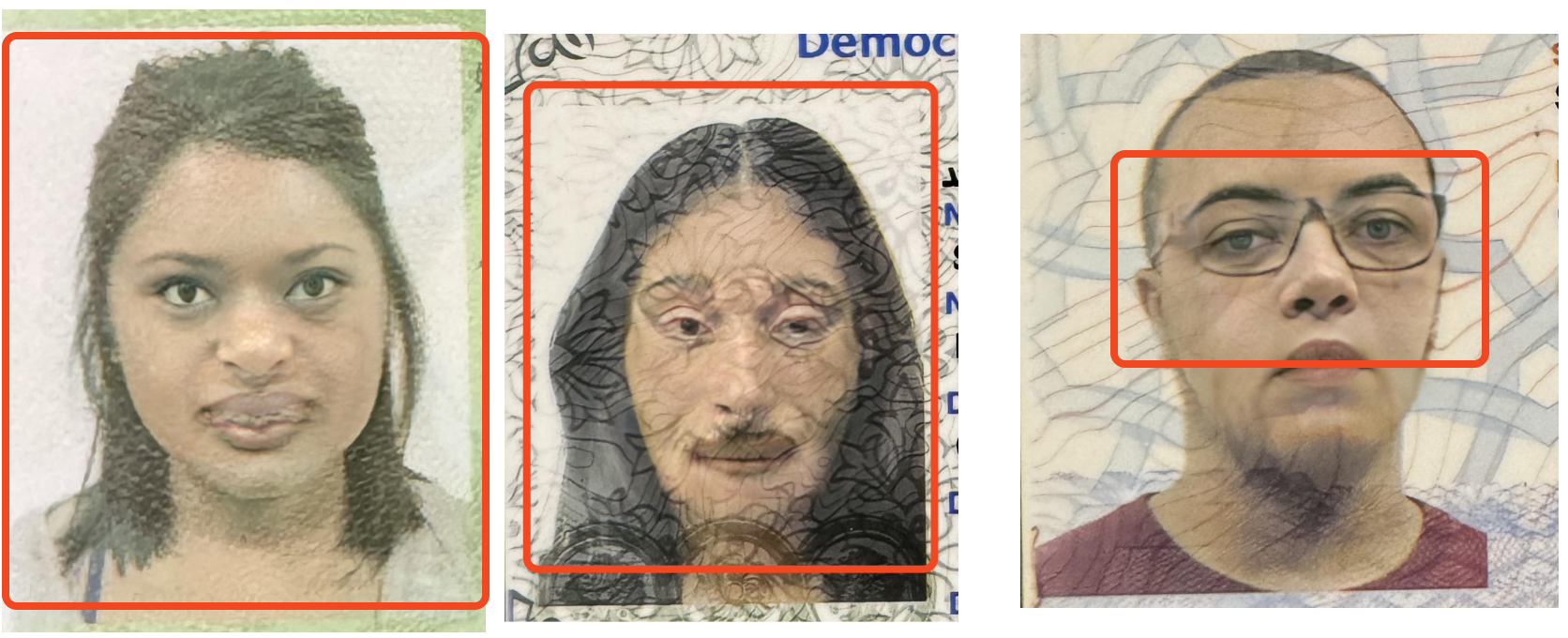}
    \caption{Face manipulation artefacts.}
    \label{fig:fantasyid_face}
\end{subfigure}
\hfill
\begin{subfigure}[t]{0.69\linewidth}
    \centering
    \includegraphics[width=\linewidth]{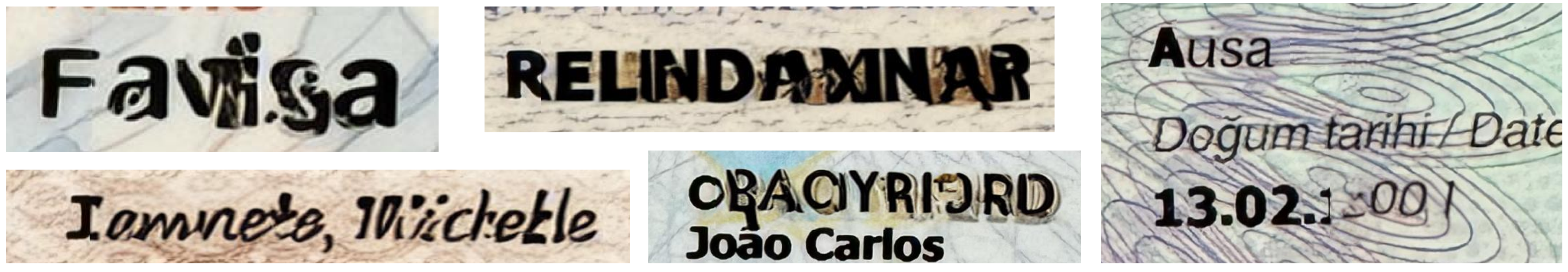}
    \caption{Text manipulation artefacts.}
    \label{fig:fantasyid_text}
\end{subfigure}

\caption{Representative GenAI-driven forgery artefacts from
FantasyID~\cite{korshunov2025fantasyid}, including font inconsistencies,
character overlap, incomplete background regeneration, and blending
artefacts.}
\label{fig:fantasyid_artifacts}
\end{figure}

The most systematic study of GenAI-assisted ID card forgery to date is
FantasyID~\cite{korshunov2025fantasyid}, which examines two principal
manipulation classes: face-region replacement and text-field editing. In
unconstrained natural images, modern generative editing systems can
produce visually plausible face replacement and scene-text inpainting.
Identity documents, however, impose a more constrained forensic setting:
fixed typography, regulated field positions, dense text regions, and
template-specific backgrounds leave little tolerance for generative
imprecision.

As shown in Figure~\ref{fig:fantasyid_artifacts}, the resulting forgeries
are heterogeneous. Some outputs achieve high visual realism and may evade
casual inspection, but many exhibit detectable failure modes, including
font mismatch, character overlap, incomplete regeneration of fine-grained
background textures, and blending artefacts around manipulated portraits.
These observations suggest that current generative pipelines can reduce
traditional copy-move and splicing cues, but do not yet produce uniformly
artefact-free identity document manipulations across all structural
regions.

\begin{figure}[tb]
\centering
\begin{subfigure}[t]{0.49\linewidth}
    \centering
    \includegraphics[width=\linewidth]{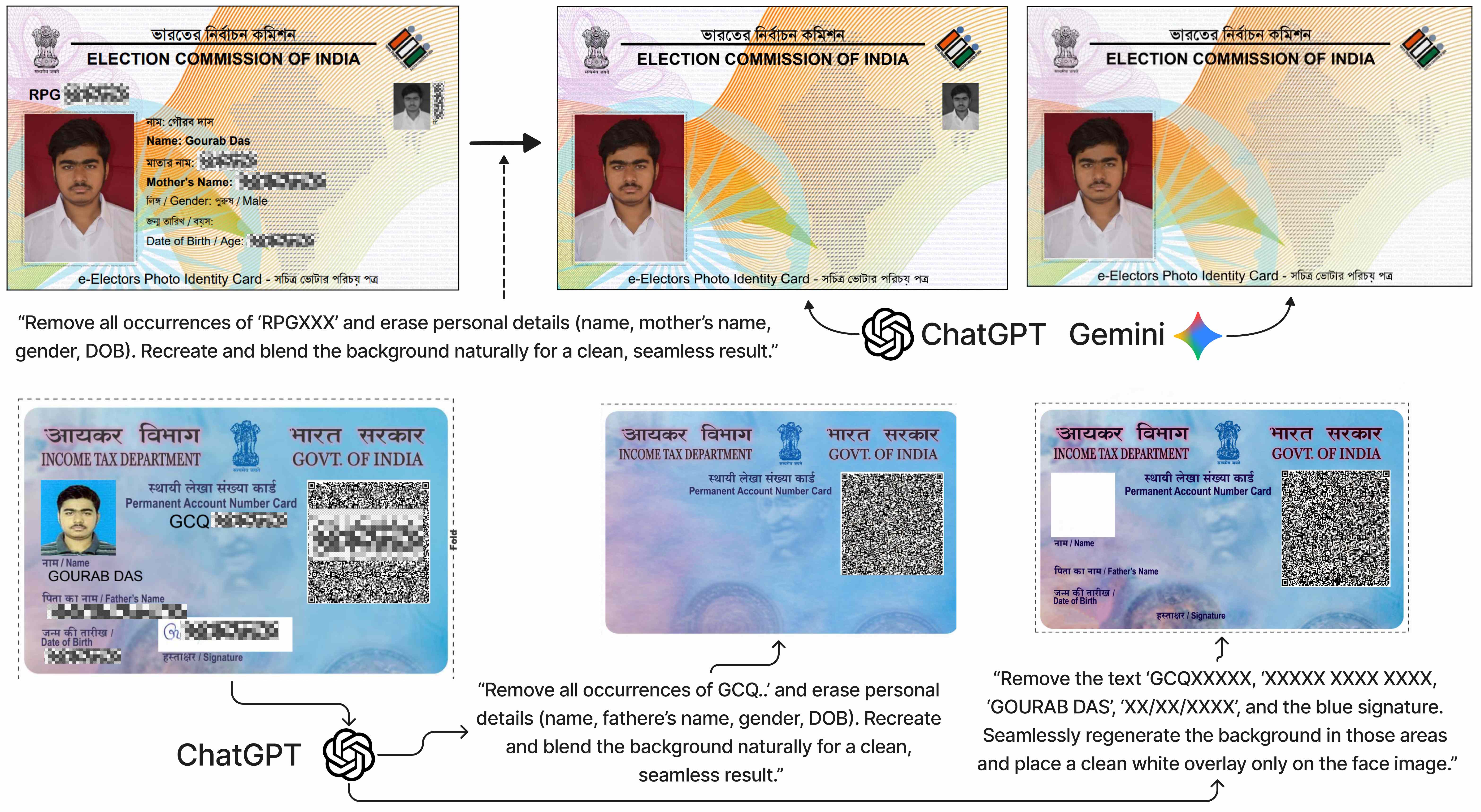}
    \caption{Extraction of a reusable blank template from a reference ID
    card image.}
    \label{fig:template_extract}
\end{subfigure}
\hfill
\begin{subfigure}[t]{0.49\linewidth}
    \centering
    \includegraphics[width=\linewidth]{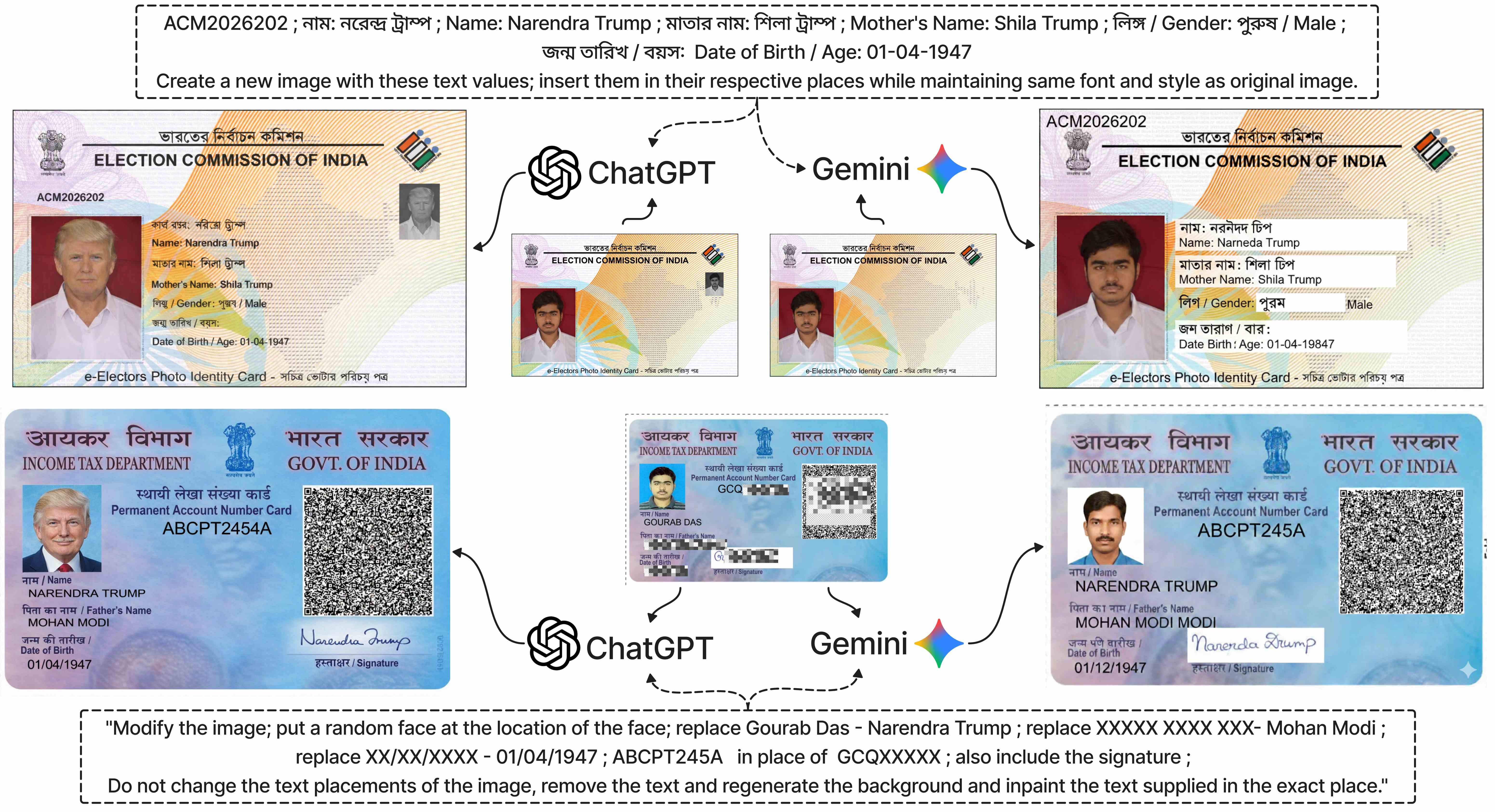}
    \caption{Generation of a modified ID card using the extracted
    template and field-level editing.}
    \label{fig:card_generate}
\end{subfigure}

\caption{Template extraction and AI-assisted ID card generation workflow,
shown for defensive analysis of generative attack capability.}
\label{fig:genai_workflow}
\end{figure}

\subsection{Full-Document Template Reconstruction via Large Multimodal Models}

Beyond specialised manipulation tools, large multimodal models can assist
in reconstructing blank document templates from reference images or
publicly available samples. Such systems may recover layout geometry,
field positions, background structure, and approximate visual styling,
especially when document templates are available online. Figure
~\ref{fig:template_extract} illustrates the extraction of a reusable blank
template from a reference card image.

Once a clean template is obtained, facial images and textual fields can
be replaced using generative editing or conventional image-editing tools,
producing a forged document that preserves the geometry and layout of the
reference template (Figure~\ref{fig:card_generate}). This workflow is
analysed here to clarify defensive requirements rather than to provide an
operational forgery recipe.

\begin{figure}[tb]
\centering
\includegraphics[width=\linewidth]{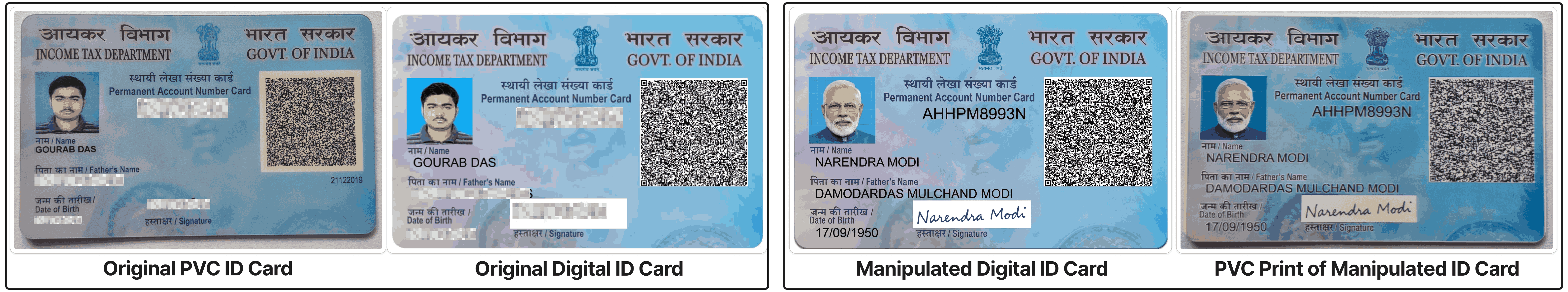}
\caption{Comparison of an original PVC card, a digital card image, a
digitally manipulated card image, and a PVC print of the manipulated
version.}
\label{fig:card-compare}
\end{figure}

The implications are particularly important for documents issued on
standard PVC substrates without advanced security features such as
holograms, optically variable devices, or specialised inks. If a digitally
forged template is reprinted on a blank PVC substrate, coarse layout and
surface appearance may remain plausible. Detection strategies relying
primarily on layout deviation or simple print-quality cues may therefore
be insufficient. Figure~\ref{fig:card-compare} shows an illustrative
comparison.

\subsection{Script-Dependent Limitations in Multimodal Generative Models}

\begin{figure}[tb]
\centering
\begin{subfigure}{0.48\linewidth}
    \centering
    \includegraphics[width=\linewidth]{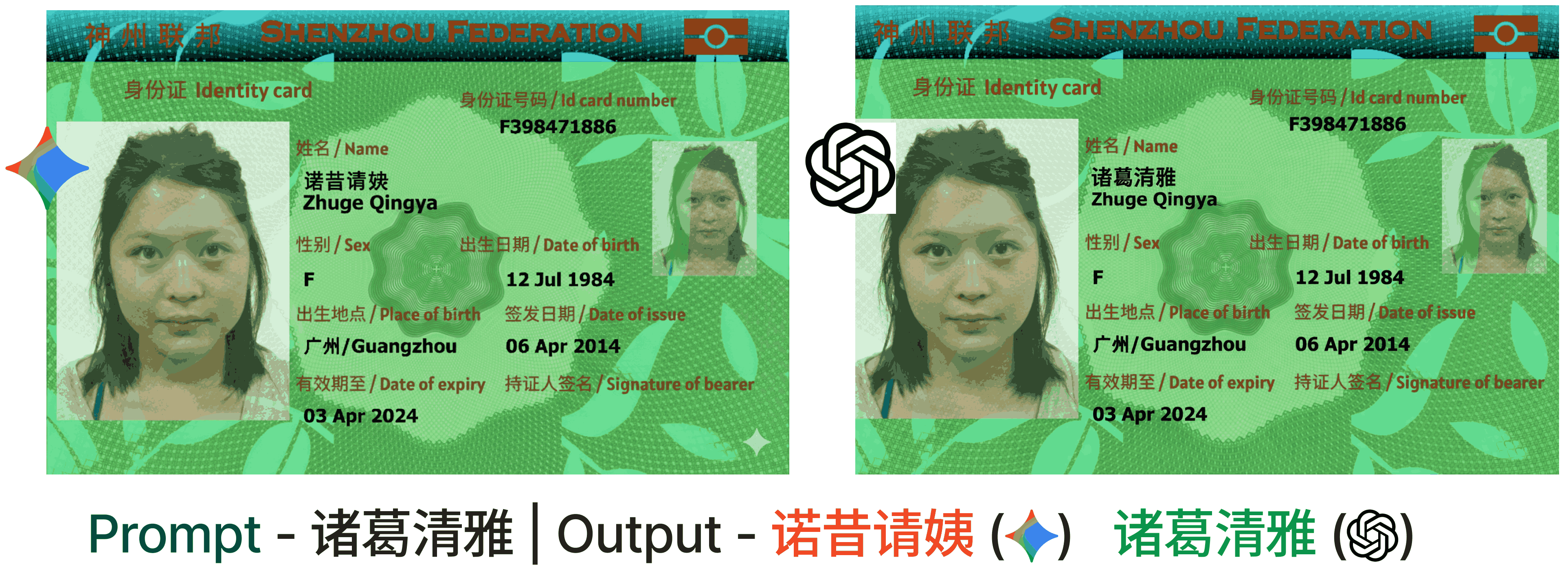}
    \caption{Chinese text inpainting. English text is reproduced plausibly, while Chinese characters are incorrect.}
    \label{fig:genai_chinese}
\end{subfigure}
\hfill
\begin{subfigure}{0.48\linewidth}
    \centering
    \includegraphics[width=\linewidth]{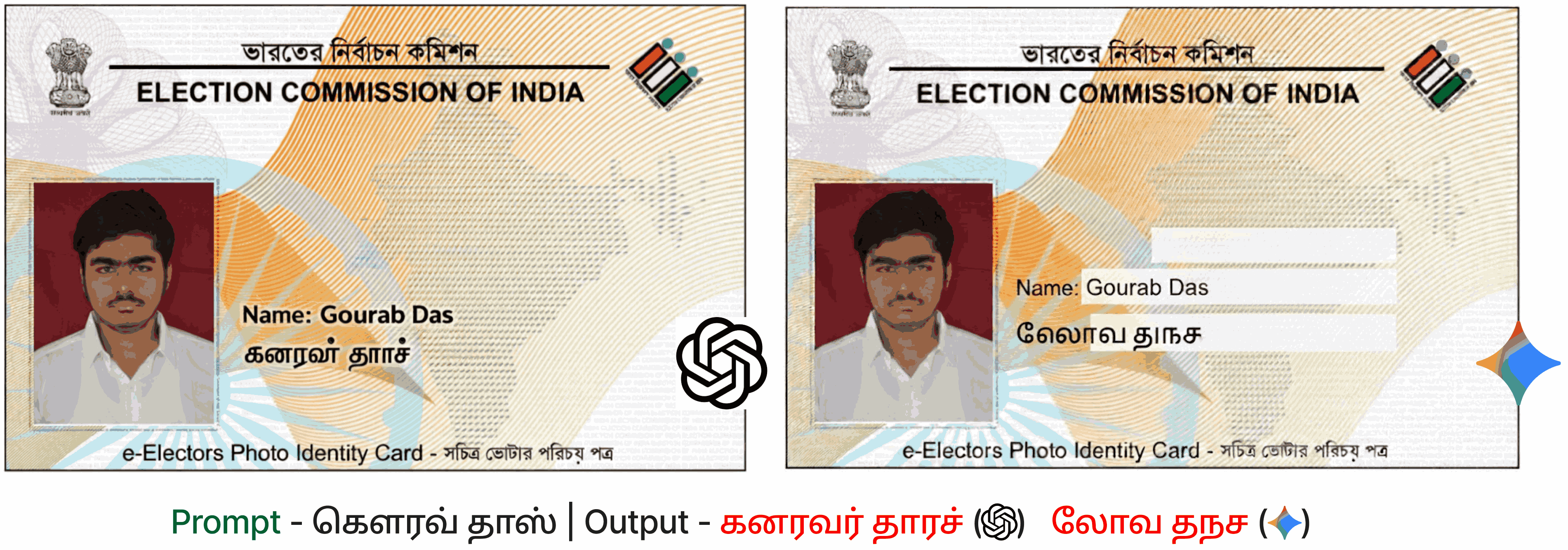}
    \caption{Tamil text inpainting showing malformed and structurally inconsistent glyphs.}
    \label{fig:genai_tamil}
\end{subfigure}
\vspace{0.8em}
\begin{subfigure}{0.995\linewidth}
    \centering
    \includegraphics[width=\linewidth]{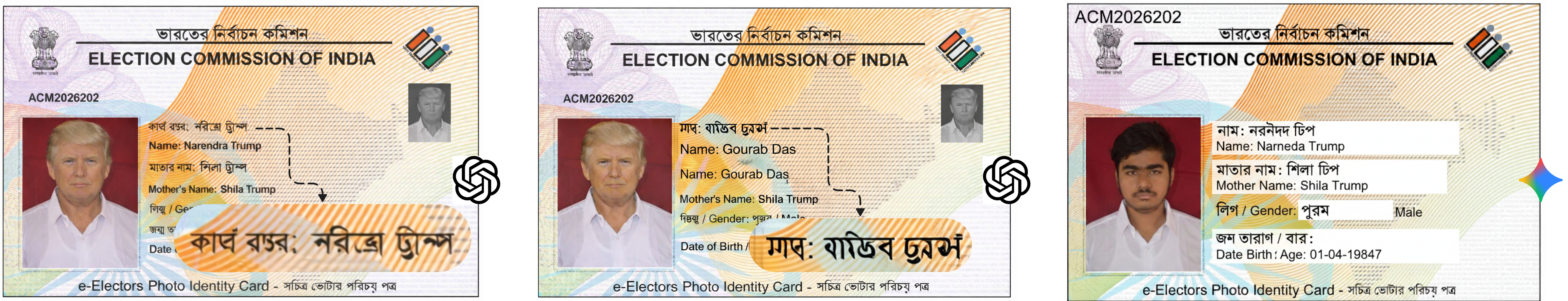}
    \caption{Regional-script examples showing character overlap, incomplete stroke formation, and typographic distortion in constrained ID fields.}
    \label{fig:genai_overlap}
\end{subfigure}
\caption{Script-dependent limitations of multimodal generative models in ID card text manipulation. Non-Latin scripts exhibit hallucinated characters, malformed glyphs, and structural inconsistencies more frequently than English-language fields.}
\label{fig:genai_comparison}
\end{figure}

Multimodal generative models often perform more reliably on English-language text inpainting than on non-Latin scripts. As illustrated in Figure~\ref{fig:genai_comparison}, script-dependent inconsistencies appear across multiple writing systems. In Figure~\ref{fig:genai_chinese}, English text is reproduced plausibly, while the corresponding Chinese field contains incorrect or hallucinated characters. In Figure~\ref{fig:genai_tamil}, Tamil text exhibits malformed or structurally inconsistent glyphs. Figure~\ref{fig:genai_overlap} shows a broader pattern in which constrained identity document layouts exacerbate typographic failures, including character overlap, incomplete stroke formation, and distortion of script-specific structural rules. Failures are particularly pronounced for complex personal names, where multi-component character sequences and script-specific ligature rules compound the generative model's typographic uncertainty.

This pattern is termed \textbf{Script-Dependent Generative Instability (SDGI)}: degradation in typographic fidelity when multimodal generative models perform text inpainting on non-Latin scripts under constrained document layouts. In a controlled evaluation of 200 inpainting attempts across Bengali, Chinese, and English script fields, script-consistent glyph reproduction was achieved in 98\% of English fields, compared with 12\% for Chinese and 6\% for Bengali. SDGI is forensically significant. Glyph validity, stroke-level consistency, inter-character spacing, and script-aware layout rules constitute detection signals specific to identity documents that are absent in generic image-forgery settings.

Regarding provenance mechanisms, some models embed visible watermarks or imperceptible C2PA metadata. In operational verification settings, such signals cannot be assumed to provide reliable protection because they may be removed, altered, lost during format conversion, or ignored by downstream systems.

\subsection{Forensic Artefacts and Detection Implications}

The detectability of GenAI-assisted forgery depends strongly on the attack
pathway. When a digitally manipulated ID card image is reprinted and
recaptured, the process introduces printer noise, substrate texture,
illumination variation, and compression artefacts. These cues remain
partly exploitable by conventional presentation attack detection
mechanisms~\cite{benalcazar2023synthetic,gonzalez2025forged}.

A more challenging scenario arises under direct injection, where a
manipulated document image is submitted into the verification pipeline
without physical recapture. In this setting, no substrate-induced artefacts
are introduced, and PAD systems based on texture irregularities or
recapture traces offer limited resistance. This threat model, formalised
in the DeepID 2025 challenge~\cite{korshunov2025deepid}, requires
detectors to reason beyond physical-domain forensic cues.

Even under injection conditions, generative pipelines may leave detectable
traces. Observed artefacts include noise-pattern inconsistencies between
authentic and synthesised regions, boundary-level blurring from blending,
compression discontinuities, and structural inconsistencies within
constrained text fields. Methods such as Noiseprint++~\cite{guillaro2023trufor}
and document-adapted manipulation detectors (Section~\ref{sec:detection})
can exploit some of these cues at pixel or region level.

Text-level forensic analysis provides an additional detection pathway.
Systematic deviations in font metrics, inter-character spacing,
stroke-width consistency, glyph validity, and script-specific typographic
rules introduce structured anomalies that are particularly relevant to
identity documents. Incorporating character-aware and layout-sensitive
analysis into PAD frameworks may therefore improve robustness against
generative injection attacks, especially for documents containing
non-Latin scripts.

\section{Deployment, Governance, and Responsible Use}
\label{sec:ethics}

\subsection{Practical Deployment Constraints}

Performance reported under controlled benchmark conditions does not
necessarily translate to operational identity verification. Public
datasets remain limited in document diversity, geographic coverage,
capture conditions, and attack realism, leaving models underexposed to the
global variability of identity document layouts, substrates, security
features, and imaging pipelines~\cite{bulatovich2022midv,polevoy2022document}.
Cross-dataset evaluations and PAD-ID-card competitions further show that
cross-country and cross-template generalisation remains a persistent
deployment challenge~\cite{tapia2024first,tapia2025second}.

Operational environments introduce additional distribution shifts,
including device variability, compression, illumination change, motion
blur, perspective distortion, and mobile onboarding artefacts. Synthetic
augmentation can improve robustness~\cite{benalcazar2023synthetic,
markham2024open}, but synthetic templates do not fully reproduce the
optical, physical, and procedural complexity of genuine circulating
documents. Moreover, research benchmarks typically define attack classes
in advance, whereas real adversaries adapt to deployed countermeasures.

A further challenge is transparency. Several state-of-the-art results are
reported on large private datasets~\cite{ruiz2025identity}, which are not
available for independent verification. This limits reproducibility,
obscures failure modes, and sustains a structural gap between academic
validation and deployment reliability.

\subsection{Privacy, Regulation, and Governance}

Robust identity document PAD requires access to genuine bona fide
documents, yet these data are highly sensitive and legally protected.
Regulations such as the General Data Protection Regulation
(GDPR)~\cite{gdpr2016} constrain the collection, storage, processing, and
sharing of identity-related information. Public release of genuine ID card
images is therefore legally and ethically restricted in many settings.

Cross-border data transfer rules, institutional review requirements, and
cloud-processing constraints further complicate collaborative training and
benchmarking. These pressures have motivated synthetic datasets
~\cite{benalcazar2023synthetic,xie2024idnet} and privacy-preserving
approaches such as the patch-based FakeIDet framework
~\cite{munoz2025fakeidet}. However, privacy-preserving design often
reduces forensic context: patch-level data protect identity information
but remove global layout, cross-field consistency, and document-level
evidence. Future benchmarks therefore require governance mechanisms that
balance privacy with forensic realism, such as controlled-access data
enclaves, standardised audit logs, anonymised evaluation protocols, and
clear restrictions on redistribution.

\subsection{Ethical and Adversarial Risks}

PAD systems deployed in financial, governmental, and border-control
settings carry substantial ethical consequences. Incorrect fraud decisions
may deny services, trigger manual investigation, or create burdens for
legitimate users. Aggregate metrics such as APCER, BPCER, and
EER~\cite{ISO30107-3-2023} are necessary but insufficient: they do not
explain individual decisions, identify affected subpopulations, or provide
case-specific evidence suitable for human review.

Bias is a related risk. Limited geographic, linguistic, or demographic
coverage in training data can produce systematic errors for underrepresented
document types, scripts, or issuing regions, with failures becoming visible
only after deployment~\cite{tapia2024first,tapia2025second}. Operational
systems should therefore include monitoring for subgroup performance,
human-review pathways, appeal mechanisms, and documentation of known
limitations.

Finally, the same generative techniques used to study document forgery can
also lower the barrier to abuse. Research on synthetic document generation
and manipulation is valuable for defensive benchmarking, but it should be
reported with care: sensitive prompts, operational forgery recipes, and
high-resolution reusable templates should not be released without
appropriate safeguards. Responsible progress in this area requires
balancing reproducibility with misuse prevention, while prioritising
detectors, benchmarks, and governance practices that improve security
without exposing unnecessary attack instructions.

\section{Future Directions}
\label{sec:future}

\subsection{Next-Generation ID-Centric Benchmarks and Datasets}

The most urgent open challenge is the absence of benchmarks that reflect
real deployment conditions. Existing public datasets are dominated by
mock-ups, synthetic templates, or geographically restricted document
types~\cite{bulatovich2022midv,polevoy2022document,xie2024idnet}, and
therefore do not capture the diversity of identity documents in global
circulation. Future datasets should include genuine bona fide samples
from multiple countries, document families, substrates, capture devices,
and linguistic contexts, while satisfying privacy and governance
constraints. Multilingual coverage is particularly important: documents
containing non-Latin scripts are necessary for evaluating semantic
consistency checks and character-level forensic analysis, especially in
view of the Script-Dependent Generative Instability identified in
Section~\ref{sec:genai}.

Attack coverage must also expand beyond the print-and-screen paradigm.
Benchmarks should jointly cover physical presentation attacks, digital
composite forgeries, injection attacks, and GenAI-driven full-document
synthesis~\cite{korshunov2025fantasyid}. Equally important is the
standardisation of localisation annotations: pixel- or region-level
ground truth would allow detection systems to be evaluated not only by
APCER/BPCER, but also by whether they identify the manipulated evidence
correctly. General localisation metrics such as mean Intersection over
Union (mIoU)~\cite{everingham2010pascal} should therefore be adapted to
identity document forensics alongside PAD metrics~\cite{ISO30107-3-2023}.
Finally, the field needs metrics for synthetic data quality that measure
forensic utility rather than only perceptual realism, since visually
plausible samples do not necessarily improve detector robustness
~\cite{markham2024open}.

\subsection{Visual and Image-Embedded Cryptographic Trust Signals}

A complementary direction is the integration of cryptographic trust
signals into the document image itself. Digitally signed QR codes, as used
in systems such as Aadhaar\footnote{\url{https://uidai.gov.in/en/ecosystem/authentication-devices-devices/qr-code-reader.html}},
allow printed demographic information to be checked against a
cryptographically protected payload without requiring online database
access. Any alteration to visible fields can then be detected through a
mismatch between the printed content and the signed data. Related
mechanisms, including content-coupled digital watermarking,
steganographic metadata, and visually verifiable signatures, deserve
further study as document-level trust anchors. Such mechanisms are
especially relevant for moderate-security ID cards that lack advanced
physical security features, because they raise the cost of both digital
and physical forgery.

\subsection{Robust Multi-Layer ID Verification Systems}

Generative AI makes single-cue detection increasingly fragile. Future
verification systems should combine physical PAD~\cite{gonzalez2025forged},
digital injection detection~\cite{korshunov2025fantasyid}, cryptographic
payload verification, template compliance checks, semantic consistency
across demographic fields, and multilingual text analysis. Rather than
treating these signals independently, robust systems should fuse them at
score, feature, or decision level so that failure of one cue does not
invalidate the entire verification pipeline. Foundation models and
vision--language architectures may provide useful representations for
such multi-evidence reasoning~\cite{oquab2023dinov2,vidit2025detecting},
but they must be adapted to document-specific forensic signals rather
than used as standalone authenticity judges.

\subsection{Explainable and Legally Accountable Forensic Models}

As PAD systems enter financial, governmental, and border-control
workflows, interpretability becomes a deployment requirement rather than
a convenience~\cite{abdullakutty2021review,shaheed2024deep}. A model that
labels a document as forged without identifying the supporting evidence
cannot support regulatory audit, human review, or legal challenge. Future
explainability methods should therefore move beyond generic saliency maps
toward document-forensic explanations: highlighting noise discontinuities
at inpainting boundaries, font metric inconsistencies in text fields,
blending artefacts in facial regions, compression anomalies, or violations
of template geometry~\cite{li2023two,zhao2021deep}. Such explanations
would improve auditability and provide analysts with actionable evidence
for correcting systematic failure modes.

\subsection{Data-Efficient and Privacy-Preserving Learning Frameworks}

Privacy constraints require learning frameworks that reduce dependence on
large centralised collections of genuine identity documents. Federated
learning~\cite{mcmahan2017communication,kairouz2021advances} offers one
path by allowing banks, government agencies, and verification providers to
train shared PAD models without exchanging raw document images. Few-shot
learning~\cite{rocamora2026simulation,sanchez2024few} provides a
complementary mechanism for rapid adaptation to new countries, templates,
and issuing authorities using only a small support set. Finally,
continual learning will be necessary for long-term deployment, since
forgery techniques evolve after detectors are released. Future systems
must incorporate new attack evidence without catastrophic forgetting,
while preserving privacy, auditability, and reproducibility.

\section{Conclusion}
\label{sec:conclusion}

This survey has traced the co-evolution of identity document forgery and
detection from opportunistic print and screen attacks to AI-assisted
manipulation and GenAI-driven document synthesis. Each increase in attack
capability has required a corresponding methodological response: from
handcrafted heuristics and global classifiers to forensic localisation,
injection-aware architectures, foundation models, and few-shot adaptation.
However, the gap between attack realism and detection robustness has
widened as generative tools have become more accessible.

The central finding is that current benchmarks and detectors remain
misaligned with operational threat models. Public datasets are dominated
by synthetic mock-ups, geographically restricted templates, and narrow
print-or-screen attack assumptions, while real deployments must handle
cross-country document variation, mobile capture artefacts, digital
injection, and generative synthesis. Our zero-shot benchmark reinforces
this Reality Gap: even the strongest publicly available models fail to
provide reliable standalone detection under security-oriented operating
points on unseen synthesised ID cards.

Three structural gaps define the next stage of research. First, the field
needs privacy-preserving benchmarks containing more realistic bona fide
documents, diverse scripts, and broader attack coverage. Second,
localisation ground truth is required so that systems can identify the
evidence supporting a forgery decision, not merely output an image-level
score. Third, evaluation protocols must jointly cover physical
presentation attacks, digital injection attacks, and GenAI-driven
synthesis. Within this broader agenda, Script-Dependent Generative
Instability (SDGI) provides a concrete example of a document-specific
forensic signal: the typographic degradation of non-Latin scripts under
generative inpainting.

Progress will require more than stronger classifiers. Robust identity
document verification must combine forensic artefact analysis,
cryptographic trust signals, semantic and layout consistency checks,
privacy-preserving learning, and continual adaptation to emerging attacks.
Generative AI has changed identity document forgery from a specialised
craft into an increasingly accessible digital capability; detection
research must now respond with equally systematic, auditable, and
document-centric methods.


\bibliographystyle{ACM-Reference-Format}
\bibliography{IDCard_Bib}

\appendix

\end{document}